\documentclass[longauth]{aa}
\usepackage[usenames,dvipsnames]{xcolor}
\usepackage{graphicx}
\graphicspath{{./figs/}}
\usepackage{txfonts}
\usepackage{gensymb}
\usepackage{tikz}
\usepackage{courier}
\usepackage{newtxtext}
\usepackage[frenchmath,varg]{newtxmath}
\usepackage{natbib,twoopt}
\usepackage[breaklinks=true,colorlinks=true,linkcolor=NavyBlue!50!Emerald,citecolor=NavyBlue!50!Emerald,filecolor=blue,urlcolor=PineGreen!50!NavyBlue]{hyperref} 
\usepackage[para,online,flushleft]{threeparttable}
\usepackage[labelfont = bf,labelsep = period]{caption} 

\bibpunct{(}{)}{;}{a}{}{,} 
\makeatletter
\newcommand{\angstrom}{\text{\normalfont\AA}}

\newcommand{\msun}{\,\hbox{$M_{\odot}$}}
\newcommand{\lsun}{\,\hbox{$L_{\odot}$}}
\newcommand{\kms}{\,\hbox{\hbox{km}\,\hbox{s}$^{-1}$}}
\newcommand{\lir}{\,\hbox{$L_{\rm IR}$}}

\newcommand{\nev}{\,\hbox{[\ion{Ne}{v}]$_{\rm 14.3 \mu m}$}}
\newcommand{\neii}{\,\hbox{[\ion{Ne}{ii}]$_{\rm 12.8 \mu m}$}}
\newcommand{\oiv}{\,\hbox{[\ion{O}{iv}]$_{\rm 25.9 \mu m}$}}
\newcommand{\oiii}{\,\hbox{[\ion{O}{iii}]$_{\rm 5007 \angstrom}$}}
\newcommand{\cii}{\,\hbox{[\ion{C}{ii}]$_{\rm 158 \micron}$}}

\newcommand{\coto}{\hbox{$\rm CO(2$-$1)$}}

\makeatletter
\newcommandtwoopt{\citeads}[3][][]{\href{http://adsabs.harvard.edu/abs/#3}%
{\def\hyper@linkstart##1##2{}%
\let\hyper@linkend\@empty\citealp[#1][#2]{#3}}}
\newcommandtwoopt{\citepads}[3][][]{\href{http://adsabs.harvard.edu/abs/#3}%
{\def\hyper@linkstart##1##2{}%
\let\hyper@linkend\@empty\citep[#1][#2]{#3}}}
\newcommandtwoopt{\citetads}[3][][]{\href{http://adsabs.harvard.edu/abs/#3}%
{\def\hyper@linkstart##1##2{}%
\let\hyper@linkend\@empty\citet[#1][#2]{#3}}}
\newcommandtwoopt{\citeyearads}[3][][]%
{\href{http://adsabs.harvard.edu/abs/#3}
{\def\hyper@linkstart##1##2{}%
\let\hyper@linkend\@empty\citeyear[#1][#2]{#3}}}
\makeatother

\DeclareSymbolFont{UPM}{U}{eur}{m}{n}
\SetSymbolFont{UPM}{bold}{U}{eur}{b}{n}
\DeclareMathSymbol{\umu}{0}{UPM}{"16}
\def\micron{\hbox{$\umu$m}}

\begin{document} 

   \titlerunning{TWIST: A Jet-Driven Outflow in ESO\,420-G13}
   \authorrunning{J.\,A. Fern\'andez-Ontiveros et al.}
   \title{A CO molecular gas wind 340\,pc away from the Seyfert 2 nucleus in ESO\,420-G13 probes an elusive radio jet}

   \author{J.\,A. Fern\'andez-Ontiveros\inst{1,2,3,4}\thanks{\email{\sf \href{mailto:j.a.fernandez.ontiveros@gmail.com}{\color{NavyBlue!50!Emerald}j.a.fernandez.ontiveros@gmail.com}, \href{mailto:juan.fernandez@inaf.it}{\color{NavyBlue!50!Emerald}juan.fernandez@inaf.it}}}, 
          K.\,M. Dasyra\inst{5,2}, 
          E. Hatziminaoglou\inst{6},
          M.\,A. Malkan\inst{7},
          \mbox{M. Pereira-Santaella\inst{8,9}},
          \mbox{M. Papachristou}\inst{5,2},
          L. Spinoglio\inst{1},
          F. Combes\inst{10},
          S. Aalto\inst{11},
          N. Nagar\inst{12},
          M. Imanishi\inst{13,14},
          P. Andreani\inst{6},
          C. Ricci\inst{15,16},
          \mbox{R. Slater\inst{17}}
          }

   \institute{
    Istituto di Astrofisica e Planetologia Spaziali (INAF--IAPS), Via Fosso del Cavaliere 100, I--00133 Roma, Italy 
    \and
    National Observatory of Athens (NOA), Institute for Astronomy, Astrophysics, Space Applications and Remote Sensing (IAASARS), GR--15236, Greece
    \and
    Instituto de Astrof\'isica de Canarias (IAC), C/V\'ia L\'actea s/n, E--38205 La Laguna, Tenerife, Spain
    \and
    Universidad de La Laguna (ULL), Dpto. Astrof\'isica, Avd. Astrof\'isico Fco. S\'anchez s/n, E--38206 La Laguna, Tenerife, Spain
    \and
    Department of Astrophysics, Astronomy \& Mechanics, Faculty of Physics, National and Kapodistrian University of Athens, Panepistimiopolis Zografou, 15784, Greece
    \and
    European Southern Observatory, Karl-Schwarzschild-Stra{\ss}e 2, D--85748, Garching, Germany
    \and
    Astronomy Division, University of California, Los Angeles, CA 90095-1547, USA
    \and
    Department of Physics, University of Oxford, Keble Road, Oxford OX1 3RH, UK
    \and
    Centro de Astrobiolog\'ia (CSIC-INTA), Ctra. de Ajalvir, Km 4, 28850, Torrej\'on de Ardoz, Madrid, Spain
    \and
    Observatoire de Paris, LERMA, College de France, CNRS, PSL Univ., Sorbonne University, UPMC, Paris, France
    \and
    Department of Space, Earth and Environment, Chalmers University of Technology, Onsala Observatory, SE-439 92 Onsala, Sweden
    \and
    Departamento de Astronom\'ia, Universidad de Concepci\'on, Concepci\'on, Chile
    \and
    National Astronomical Observatory of Japan, National Institutes of Natural Sciences (NINS), 2-21-1 Osawa, Mitaka, Tokyo 181-8588, Japan
    \and
    Department of Astronomy, School of Science, Graduate University for Advanced Studies (SOKENDAI), Mitaka, Tokyo 181-8588, Japan
    \and
    N\'ucleo de Astronom\'ia de la Facultad de Ingenier\'ia, Universidad Diego Portales, Av. Ej\'ercito Libertador 441, Santiago, Chile
    \and
    Kavli Institute for Astronomy and Astrophysics, Peking University, Beijing 100871, China
    \and
    Direcci\'on de Formaci\'on General, Facultad de Educaci\'on y Cs. Sociales, Universidad Andres Bello, Sede Concepci\'on, autopista Concepci\'on-Talcahuano 7100, Talcahuano, Chile
   }

   \date{\today} 

 
   \abstract{A prominent jet-driven outflow of CO(2--1) molecular gas is found along the kinematic minor axis of the Seyfert 2 galaxy \mbox{ESO\,420-G13}, at a distance of $340$--$600\, \rm{pc}$ from the nucleus. The wind morphology resembles a characteristic funnel shape, formed by a highly collimated filamentary emission at the base, likely tracing the jet propagation through a tenuous medium, until a bifurcation point at $440\, \rm{pc}$ where the jet hits a dense molecular core and shatters, dispersing the molecular gas into several clumps and filaments within the expansion cone. We also trace the jet in ionised gas within the inner $\lesssim 340\, \rm{pc}$ using the \neii \ line emission, where the molecular gas follows a circular rotation pattern. The wind outflow carries a mass of $\sim 8 \times 10^6\, \rm{M_\odot}$ at an average wind projected speed of $\sim 160\, \rm{\kms}$, which implies a mass outflow rate of $\sim 14\, \rm{M_\odot\,yr^{-1}}$. Based on the structure of the outflow and the budget of energy and momentum, we discard radiation pressure from the active nucleus, star formation, and supernovae as possible launching mechanisms. ESO\,420-G13 is the second case after NGC\,1377 where the presence of a previously unknown jet is revealed due to its interaction with the interstellar medium, suggesting that unknown jets in feeble radio nuclei might be more common than expected. Two possible jet-cloud configurations are discussed to explain the presence of an outflow at such distance from the AGN. The outflowing gas will likely not escape, thus a delay in the star formation rather than quenching is expected from this interaction, while the feedback effect would be confined within the central few hundred parsecs of the galaxy.}

   \maketitle

   \section{Introduction}\label{intro}

Feedback from star formation and active galactic nuclei (AGN) has been invoked to ease the tension between cosmological simulations of galaxy formation and evolution, and observations across all redshifts. Massive and fast outflows are rather common among luminous AGN, being detected in different gas phases and physical scales: from parsec-scale ultra-fast ($0.1\, \rm{c}$) outflows in X-rays to kpc-scale winds detected in atomic, molecular and ionised gas with velocities reaching or even exceeding $1000\, \rm{\kms}$ (e.g. \citeads{2017A&A...601A.143F}). By removing and/or heating the inter-stellar medium (ISM) such outflows may be able to suppress the star formation in the host galaxy. Indeed, simulations of massive galaxy formation in the past 15 years rely on the feedback effect from supermassive black holes (SMBHs) to reproduce the properties of massive galaxies observed in the local Universe (\citeads{2005ApJ...620L..79S,2006MNRAS.370..645B,2006MNRAS.365...11C,2012RAA....12..917S,2013MNRAS.433.3297D,2015MNRAS.449.4105C,2015ARA&A..53...51S,2018MNRAS.479.4056W}). From the observational point of view, however, it has yet to be established the overall impact that AGN-driven outflows cause to the star formation in their host galaxies.

Vast observational efforts in the last years have been dedicated to the detection and characterisation AGN winds in molecular gas (e.g. \citeads{2012A&A...543A..99C,2014A&A...562A..21C,2010A&A...518L.155F,2013A&A...549A..51F,2013ApJ...775..127S,2014A&A...565A..46D,2015A&A...580A..35G,2017A&A...601A.143F,2018ApJ...863..143C,2018ApJ...859..144A,2019MNRAS.483.4586F,2019ApJ...875L...8R,2019arXiv190401483F}), producing a relatively large collection of individual sources. Recently, a compilation of 45 galaxies analysed by \citetads{2019MNRAS.483.4586F} showed that the outflow mass is largely dominated by the molecular and neutral gas in AGN galaxies, revealing also that the presence of an AGN strongly boost the mass outflow rates. However, essential aspects such as the overall occurrence of outflows in AGN host galaxies and the driving mechanisms of such outflows still remain unclear. In particular, the role of jets among moderate- to low-luminosity AGN is not understood. Some examples where the presence of a jet is required to launch the observed outflows are NGC\,1266 \citepads{2011ApJ...735...88A,2013ApJ...779..173N}, NGC\,1068 \citepads{2014A&A...567A.125G}, IC\,5063 \citepads{2015A&A...580A...1M,2016A&A...595L...7D}, and NGC\,5643 \citepads{2018ApJ...859..144A}. A clearer case was found in NGC\,1377, due to the collimated morphology of the molecular gas wind \citepads{2012A&A...546A..68A}. According to theoretical predictions, jets are very efficient at delivering a large amount of energy and momentum at large distances from the active nucleus (e.g. \citeads{2012ApJ...757..136W,2012ARA&A..50..455F}). However, jet-driven outflows were identified by \citetads{2019MNRAS.483.4586F} only in a few powerful radio-loud AGN in their sample.
\begin{figure*}
\centering
\includegraphics[width = 0.8\textwidth]{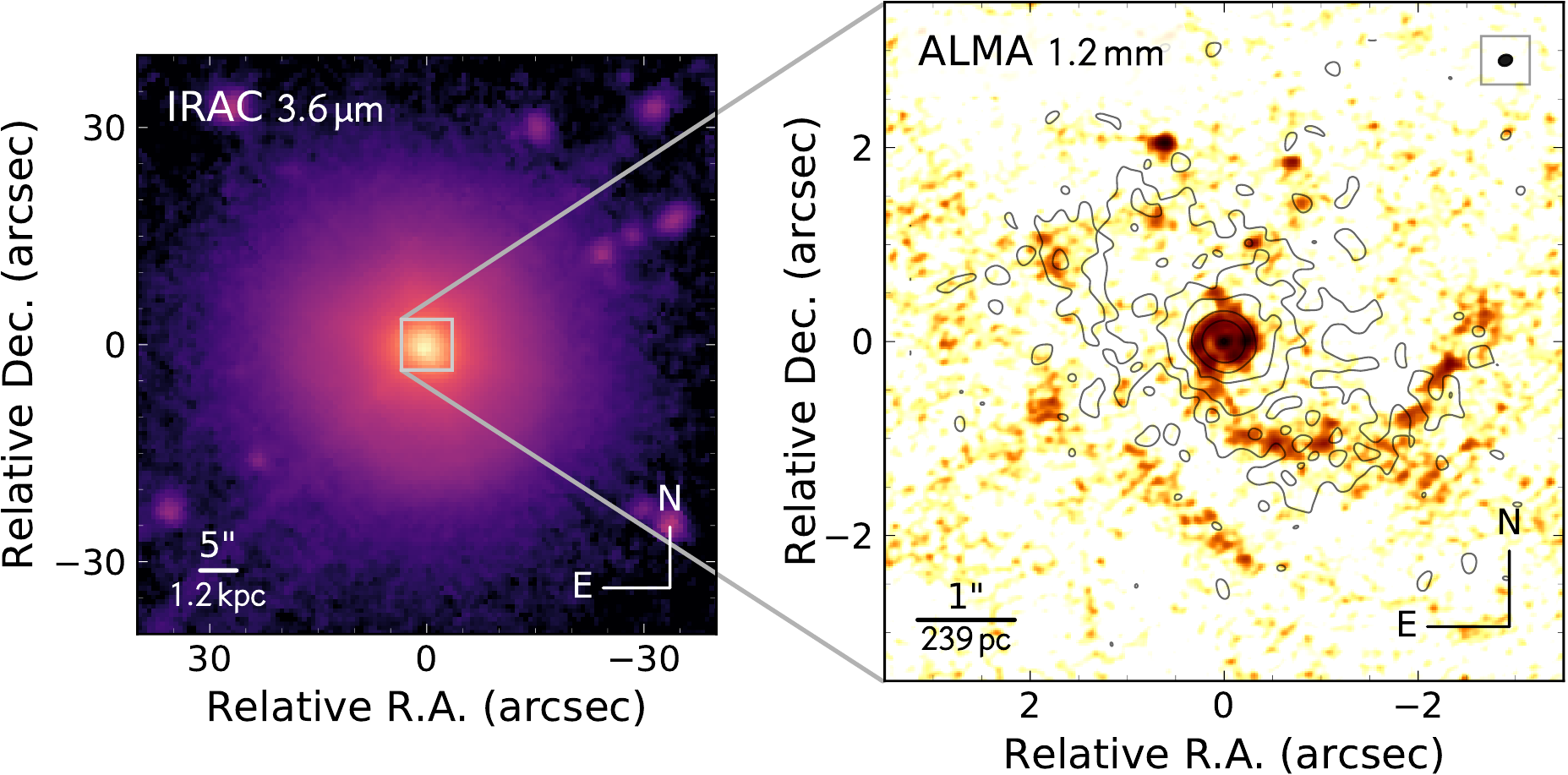}
\caption{\textit{Left:} the early-type galaxy ESO\,420-G13 imaged by \textit{Spitzer}/IRAC in the $3.6\, \rm{\micron}$ continuum. \textit{Right:} ALMA resolves the cold dust continuum at $1.2\, \rm{mm}$ into several knots defining a spiral pattern (background map). The nuclear disc is also revealed in the VLT/VISIR warm dust continuum at $12.7\, \rm{\micron}$ adjacent to the \neii \ emission line (black contours, starting from $2 \times$\,\textsc{rms} with a spacing of $\times 10^{N/3}$; \citeads{2014MNRAS.439.1648A}). The nuclear point-like source is possibly associated with synchrotron emission from the AGN, also detected at radio frequencies \citepads{1996ApJS..103...81C,1998AJ....115.1693C}. The synthesised beam size is shown in the upper-right corner.}\label{fig_zoom}
\end{figure*}

A systematic survey of outflows in a representative sample of galaxies is required to determine the occurrence of this phenomenon and to probe the overall impact of feeding and feedback in AGN, spanning a luminosity range that brackets the knee of the AGN luminosity function. While this is the aim of our ALMA ``Twelve micron sample WInd STatistics'' (TWIST) survey (see Section\,\ref{twist}), in this study we report the first outflow detected in the nucleus of the Seyfert 2 galaxy ESO\,420-G13, a Southern target (\mbox{$\alpha\,=\,\rm{04^{h}\,13^{m}\,49.7^{s}}$}, \mbox{$\delta\,=\,\rm{-32^{d}\,00^{m}\,25^{s}}$}) located at $D = 49.4\, \rm{Mpc}$\footnote{Flat-$\Lambda$CDM, $H_0 = 73\, \rm{\kms\,Mpc^{-1}}$, $\Omega_{\rm m} = 0.27$, $z = 0.01205$ (this work).}. This object has been classified as a Luminous InfraRed Galaxy (LIRG; $L_{\rm FIR} \sim 10^{11}\, \rm{L_\odot}$, \citeads{2003AJ....126.1607S}), showing a composite spectrum within the inner $1\, \rm{kpc}$ with a mixed contribution from Seyfert 2 emission and a post-starburst component, the latter characterised by an A-star dominant continuum \citepads{2017ApJS..232...11T}. On the other hand, the optical and near-infrared (near-IR) morphology is rather characteristic of an early type SA0 galaxy, with nuclear dusty spirals revealed in the mid-IR continuum \citepads{2014MNRAS.439.1648A}. For an integrated \textit{K}~band magnitude of $9.51 \pm 0.08\, \rm{mag}$ taken from the \textit{HyperLeda}\footnote{\url{http://leda.univ-lyon1.fr}} database \citepads{2014A&A...570A..13M}, we estimate a supermassive black hole mass of about $M_{\rm BH} \sim 4 \times 10^8\, \rm{M_\odot}$, using the correlation in \citetads{2013ARA&A..51..511K}. Regarding the star formation, we derive a $SFR = 6.1 \pm 0.6\, \rm{M_\odot\,yr^{-1}}$, using the PAH $11.3 \, \rm{\micron}$ flux given by \citetads{2014ApJ...790..124S} and the calibration by \citetads{2012ApJ...746..168D}, close to the $\sim 12\, \rm{M_\odot\,yr^{-1}}$ derived from the \cii \ emission \citepads{2014A&A...568A..62D,2014ApJ...790...15S}. A solid evidence for the presence of an AGN comes from the detection of high-excitation mid-IR fine-structure lines in the \textit{Spitzer}/IRS spectra of ESO\,420-G13 \citepads{2015ApJS..218...21L}, since photoionisation by stars drops abruptly after the helium ionisation edge at $54\, \rm{eV}$ and thus cannot produce a significant amount of lines such as [\ion{Ne}{v}]$_{\rm 14.3, 24.3 \mu m}$ with $97.12\, \rm{eV}$ and \oiv \ with $54.94\, \rm{eV}$ (e.g. \citeads{2005ApJ...633..706W,2006ApJ...640..204A,2009MNRAS.398.1165G}). At radio wavelengths there is bright non-resolved emission within the inner $< 17''$ \citepads{1996ApJS..103...81C,1998AJ....115.1693C}. Still, ESO\,420-G13 show a IR-to-radio ratio of $q_{\rm IR} = 2.7$ (this work; $q_{\rm IR}$ definition from \citeads{2010MNRAS.402..245I,2019MNRAS.483.4586F}), typical of starburst galaxies, while powerful radio-loud galaxies have $q_{\rm IR} \lesssim 1.8$. In the X-rays, \citetads{2018A&A...620A.140T} found a slight excess at $6.4\, \rm{keV}$, but could not confirm the presence of a Fe\,K$\alpha$ line. The faint luminosity ($L_{\rm 2-10 keV} \lesssim 2 \times 10^{40}\, \rm{erg\,s^{-1}}$), the low absorption column estimate ($N_{\rm H} \sim 6 \times 10^{21}\, \rm{cm^{-2}}$), and the steep slope of $\Gamma \sim 3$ in the $2$--$7\, \rm{keV}$ continuum range suggest that the X-ray emission might be dominated by star formation while the AGN remains obscured in this range. No CO(1--0) emission was detected above a $3 \sigma$ level of $0.7\, \rm{Jy}$ in previous single dish observations by \citetads{1996A&AS..115..439E}, which corresponds to $< 180\, \rm{Jy\,\kms}$ assuming a total line width of $250\, \rm{\kms}$ for the unresolved galaxy, or $M_{\rm H_2} \lesssim 10^9\, \rm{M_\odot}$ using the \citetads{1997ApJ...478..144S} formula with $\alpha = 0.8\, \rm{M_\odot\,(K\,\kms\,pc^2)^{-1}}$ for the ISM of active star-forming galaxies \citepads{2013ARA&A..51..207B}.

This work is organised as follows. Observations and data reduction are detailed in Section\,\ref{obs}, the properties of the detected outflow are characterised in Section\,\ref{results}. The main results are discussed in Section\,\ref{discuss}, and the final conclusions are presented in Section\,\ref{sum}.

\section{Observations}\label{obs}

\subsection{A source from the TWIST survey}\label{twist}
ESO\,420-G13 is one of the galaxies included in the Twelve micron WInd STatistics (TWIST) project (Fern\'andez-Ontiveros et al. in prep.), a CO(2--1) molecular gas survey of 41 galaxies drawn from the $12$ micron sample \citepads{1993ApJS...89....1R}. Half of the sample was acquired in $27\,\rm{h}$ of observing time with the Atacama Large Millimeter/submillimeter Array (ALMA) interferometer (PI: \mbox{M. Malkan}; Programme IDs: 2017.1.00236.S, 2018.1.00366.S), located at Llano de Chajnantor (Chile), while the other half corresponds to data available in the ALMA scientific archive. The TWIST sample covers the knee of the $12\, \rm{\micron}$ luminosity function of Seyfert galaxies in the nearby Universe ($D = 10$--$50\, \rm{Mpc}$), and thus the results obtained from TWIST will be representative of the bulk population of active galaxies. This is crucial to quantify the global AGN feeding and feedback and measure the time-averaged impact of these processes.

\subsection{ALMA millimetre interferomery}

ESO\,420-G13 was observed with ALMA on December 8th, 2017 using the 12\,m array (PI: \mbox{M. Malkan}; Programme ID: 2017.1.00236.S). The spectral setup was optimised for the CO(2--1) transition line in band 6, at $230.5380\, \rm{GHz}$ rest frequency (excitation temperature $T_{\rm ex} = 16.6\, \rm{K}$, critical density $n_{\rm crit} = 2.7 \times 10^3\, \rm{cm^{-3}}$), which was selected as an optimal mass tracer in terms of angular resolution and sensitivity. A velocity channel width of $2.4\, \rm{\kms}$ and a total bandwidth of $2467\, \rm{\kms}$ were selected for the correlator setup. Additionally, two continuum bands at $\sim 1.2\, \rm{mm}$ were acquired to trace the cold dust continuum, and a spectral band centred on the CS(5--4) at $244.9356\, \rm{GHz}$. This line (with $T_{\rm Eup} = 35.3\, \rm{K}$) traces denser gas and has a critical density typically three orders of magnitude higher than the CO(2--1) line. The data were calibrated using the Common Astronomy Software Applications package (\textsc{casa}), pipeline v5.1.1-5, and the scripts provided by the observatory. These included the flagging of four antennae due to either a bad value in the system temperature ($T_{\rm sys}$), amplitude outliers, or a spike in $T_{\rm sys}$. Imaging and post-processing were done using our own scripts under \textsc{casa} v5.4.0-70. The image reconstruction was performed using the standard \texttt{hogbom} deconvolution algorithm with \texttt{briggs} weighting and a robustness value of $2.0$. This is equivalent to using natural weighting for the image reconstruction, and allows us to recover extended flux that might be filtered for lower robustness values, which give a larger weight to the most extended baselines. The beam size in the CO(2--1) spectral window is $0\farcs11 \times 0\farcs14$ at a position angle of PA\,$= 108\fdg5$, which corresponds to $26.3 \times 33.5\, \rm{pc^2}$ at a distance of $49.4\, \rm{Mpc}$. The field-of-view (FOV) has a diameter of $27''$ ($\approx 6\, \rm{kpc}$) and was covered using a single pointing. The largest angular scale that could be resolved in this antennae configuration is $0\farcs8$ ($\sim 190\, \rm{pc}$). Average continuum images were also obtained for each of the two frequency side-bands, discarding those channels with either CO(2--1) or CS(5--4) emission line detection. The masking procedure for the continuum data was run interactively during the cleaning process. Additionally, a deeper continuum image at $\sim 240\, \rm{GHz}$ ($1.2 \, \rm{mm}$) was obtained by combining all the continuum channels within the four spectral windows (left panel in Fig.\,\ref{fig_zoom} and upper left panel in Fig.\,\ref{fig_moments}). The spectral datacubes of the emission lines were produced with a channel with of $\sim 10\, \rm{\kms}$ and a pixel size of $0\farcs02$. The emission line regions were automatically masked during the cleaning process in the spectral cubes using the ``auto-multithresh'' algorithm in \texttt{tclean}. The continuum emission was then subtracted in the spatial frequency domain --\,i.e. prior to the image reconstruction\,-- using a zero degree polynomial between the adjacent continuum channels at both sides of the respective emission lines. Finally, the datacubes were corrected for the attenuation pattern of the primary beam. The rms sensitivity of the processed cubes is $0.02\, \rm{mJy\,beam^{-1}}$ in continuum and $0.5\, \rm{mJy\,beam^{-1}}$ for a $10\, \rm{\kms}$ line width.

\subsection{Mid-infrared narrow-band imaging}

ESO\,420-G13 was imaged using the VISIR\footnote{\textsc{Vlt} spectrometer and imager for the mid-infrared \citepads{2004Msngr.117...12L}.} instrument, installed at the Very Large Telescope on Cerro Paranal (Chile) during the night of January 19th, 2006 (programme ID: 076.B-0696). These observations include two narrow-band filters in the \textit{N}~band, NEII ($\lambda_c = 12.81\, \rm{\micron}$, $\Delta \lambda = 0.21\, \rm{\micron}$) and NEII\_2 ($\lambda_c = 13.04\, \rm{\micron}$ $\Delta \lambda = 0.22\, \rm{\micron}$), which were acquired using a chopping throw of $10''$ and a pixel scale of $0\farcs127$. The final reduced images, processed using the VISIR pipeline delivered by ESO\footnote{European Southern Observatory.} as described in \citetads{2014MNRAS.439.1648A}, were taken directly from the SubArcSecond Mid-InfRared Atlas of Local AGN (SASMIRALA\footnote{\url{http://dc.g-vo.org/sasmirala}}). At a redshift of $z = 0.01205$ the NEII\_2 filter contains the \neii \ fine-structure line redshifted to $12.96\, \rm{\micron}$, while the adjacent NEII filter provides a measurement of the $\sim 12.5$--$12.7\, \rm{\micron}$ mid-infrared (mid-IR) continuum in the rest frame. Due to the similar width and transmission of the two narrow-band filters, we directly subtracted the continuum image from the line image. The difference in transmission between the two filters is of the order of the photometric error, i.e. about $5\%$ as measured for the associated calibrators in the SASMIRALA atlas. Due to the lack of reference stars in the VISIR FOV, a reliable astrometric calibration for the mid-IR images could not be obtained. Therefore, we assume the position of the bright mid-IR nucleus to be coincident with that of the compact nuclear continuum source in the ALMA $1.2\, \rm{mm}$ map.


\section{Results}\label{results}

\begin{figure*}
\centering
\includegraphics[width=0.5\textwidth]{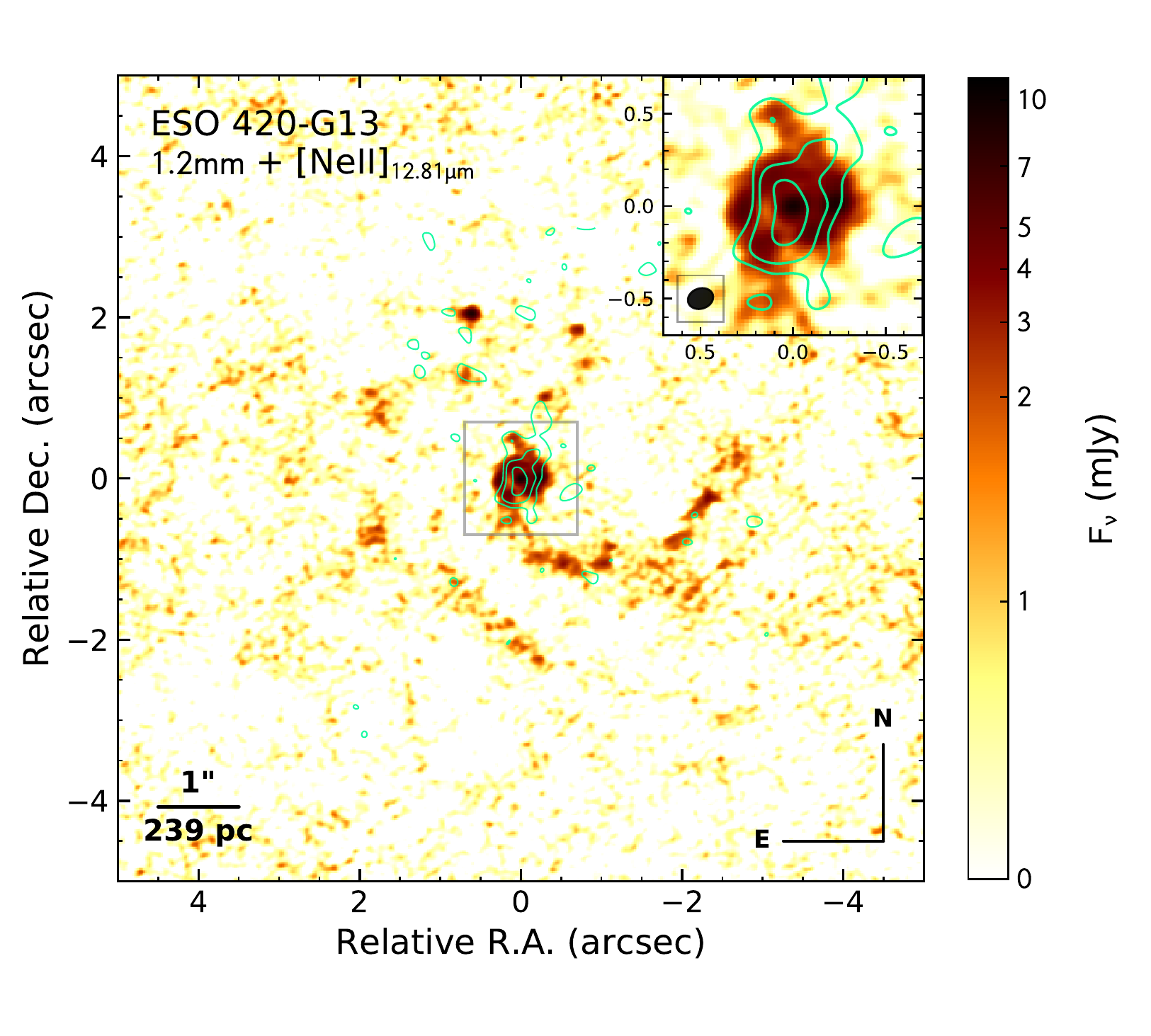}~
\includegraphics[width=0.5\textwidth]{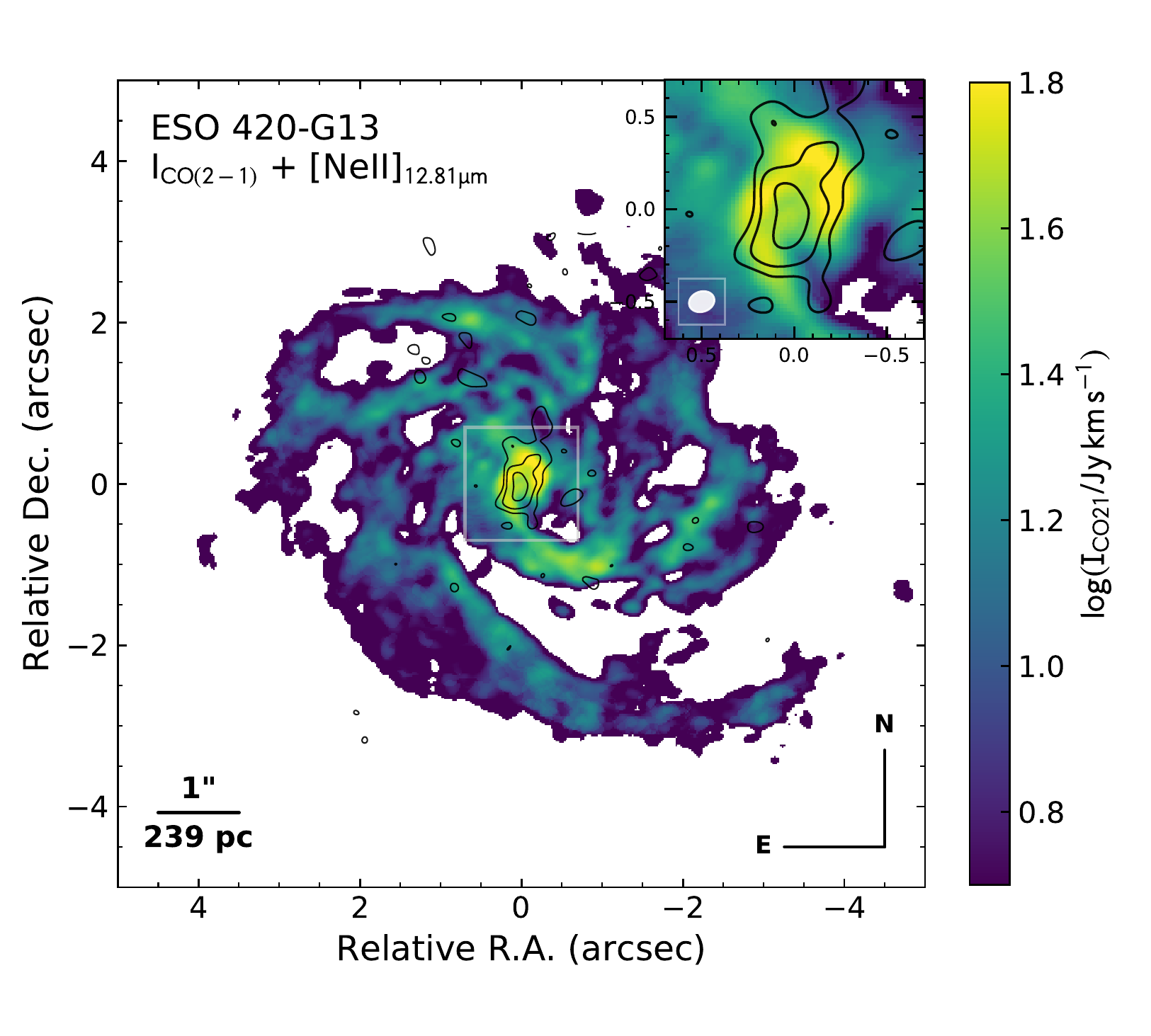}
\includegraphics[width=0.5\textwidth]{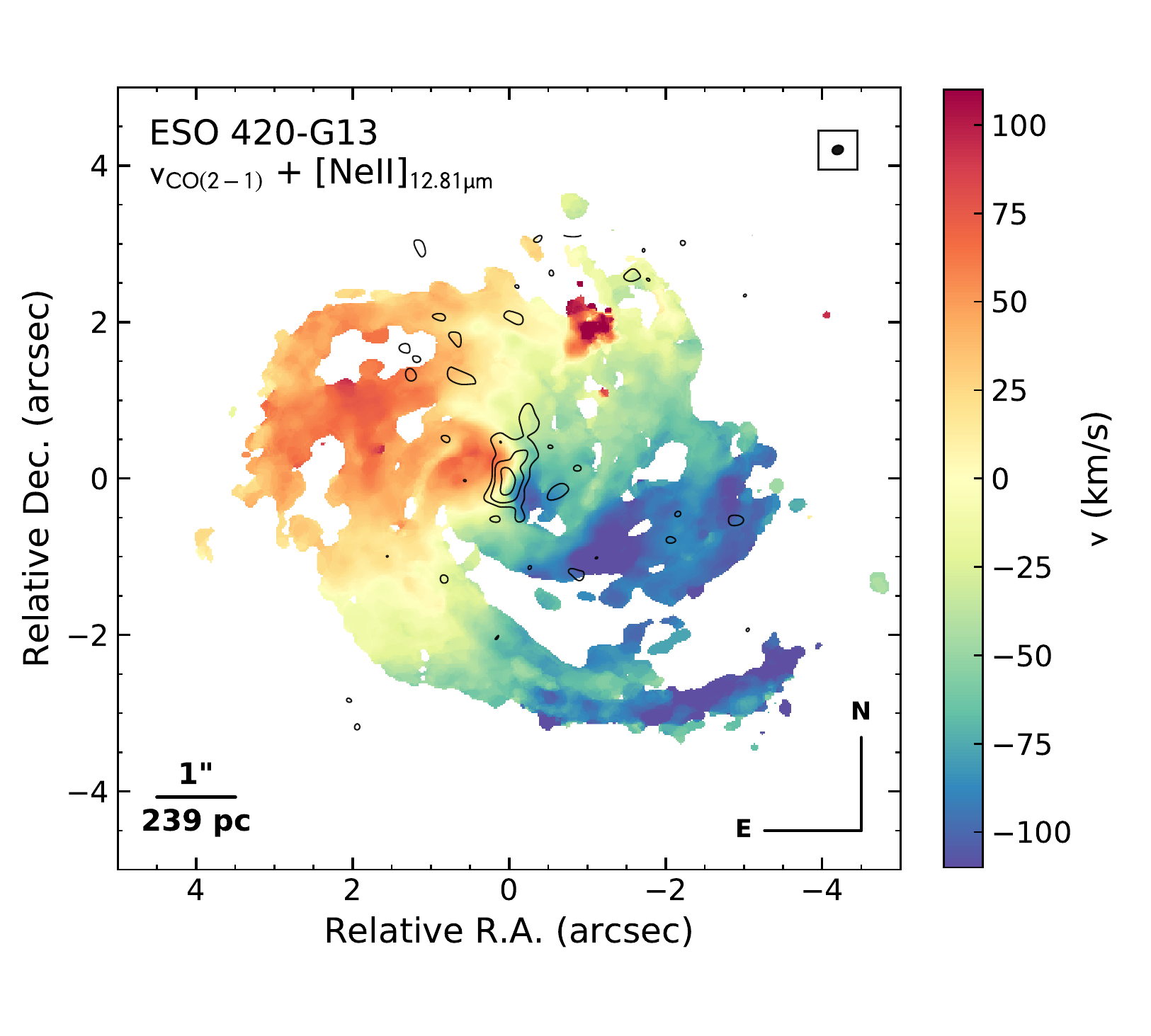}~
\includegraphics[width=0.5\textwidth]{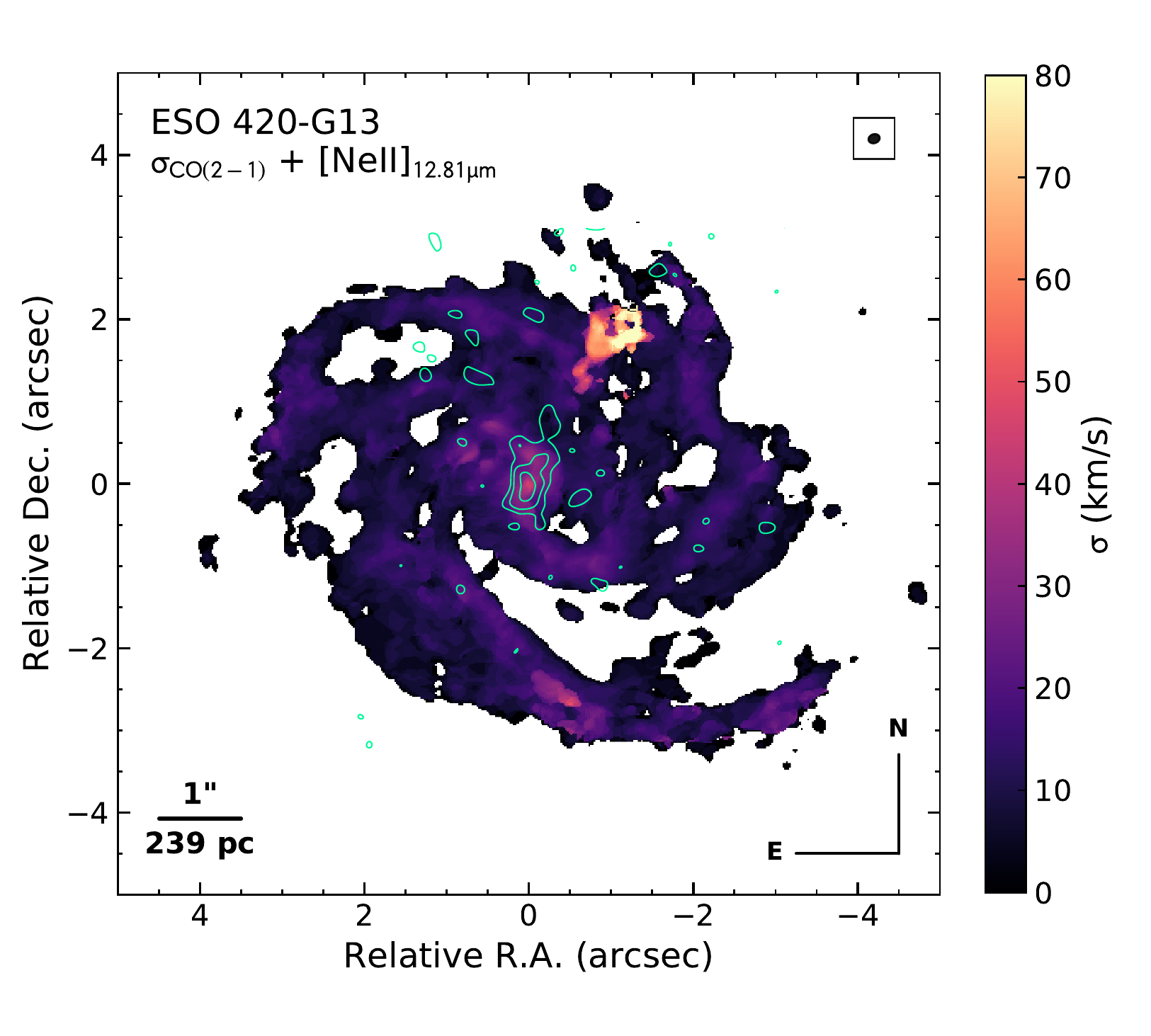}
\caption{ALMA maps for the $1.2\, \rm{mm}$ continuum (upper left), CO(2--1) intensity (upper right), CO(2--1) average velocity (lower left), and CO(2--1) average velocity dispersion (lower right) maps for the Seyfert 2 galaxy ESO\,420-G13. The zoomed inset panels correspond to the grey square regions indicated in the corresponding maps. Contours in all panels correspond to the \neii \ line emission from VLT/VISIR (starting from $3 \times$\,\textsc{rms} with a spacing of $\times 10^{N/4}$). Assuming trailing spiral arms implies that the South-East extreme of the kinematic minor axis is the nearest point to us. The synthesised beam size is shown in the inset panels and in the upper-right corner of the lower maps.}\label{fig_moments}
\end{figure*}

At first sight ESO\,420-G13 resembles an early type galaxy in the \textit{Spitzer}/IRAC $3.6\, \rm{\micron}$ image (left panel in Fig.\,\ref{fig_zoom}) with an inner kpc-size disc revealed by both warm and cold dust emission (right panel). The superior angular resolution of ALMA allows us to resolve the dust emission at $1.2\, \rm{mm}$ into three main components: a circumnuclear ring with $\sim 0\farcs8$ diameter ($190\, \rm{pc}$), an unresolved point-like source in the centre of this ring, and a kpc-size spiral structure. The latter is more prominent in the CO(2--1) intensity map (upper-right panel in Fig.\,\ref{fig_moments}), where the cold molecular gas can be traced at radii up to $\sim 1.1\, \rm{kpc}$ away from the nucleus. The inner ring is presumably located at the inner Lindblad resonance, showing bright CO(2--1) emission possibly associated with active star formation. This is also the case for the bright $1.2\, \rm{mm}$ continuum spot on the North-East, which also has enhanced molecular gas emission. Inside the ring we detect diffuse line emission, however the lack of a bright knot in CO(2--1) is in contrast with the bright core found in the $1.2\, \rm{mm}$ continuum at the centre of the ring ($\sim 0.30 \pm 0.08\, \rm{mJy}$). This suggests that the millimetre radio emission from the nucleus could be non-thermal synchrotron radiation associated with the active nucleus. In this regard, a bright compact radio core was previously detected at $1.4\, \rm{GHz}$ by \citetads{1996ApJS..103...81C}, albeit at lower angular resolution ($17''$). Further radio observations at high-angular resolution would be required to probe the inner jet morphology and the synchrotron nature of the emission. The total CO(2--1) intensity integrated in the ALMA map is $260 \pm 2\, \rm{Jy\,\kms}$, which corresponds to a mass of $M_{\rm tot} = (3.07 \pm 0.02) \times 10^8\, \rm{M_\odot}$, assuming optically-thick gas and $\alpha = 0.8\, \rm{M_\odot\,(K\,\kms\,pc^2)^{-1}}$, as in LIRG and ULIRG galaxies \citepads{2013ARA&A..51..207B}. This is in agreement with the upper limit of $< 10^9\, \rm{M_\odot}$ derived from previous single-dish observations \citepads{1996A&AS..115..439E}. Faint CS(5--4) emission is detected in the nucleus ($1.3 \pm 0.2\, \rm{Jy\,\kms}$) and tentatively also in two molecular gas clumps in the spiral arms with integrated fluxes in the $0.3$--$0.5\, \rm{Jy\,\kms}$ range.

\begin{figure*}
\centering
\includegraphics[width = \textwidth]{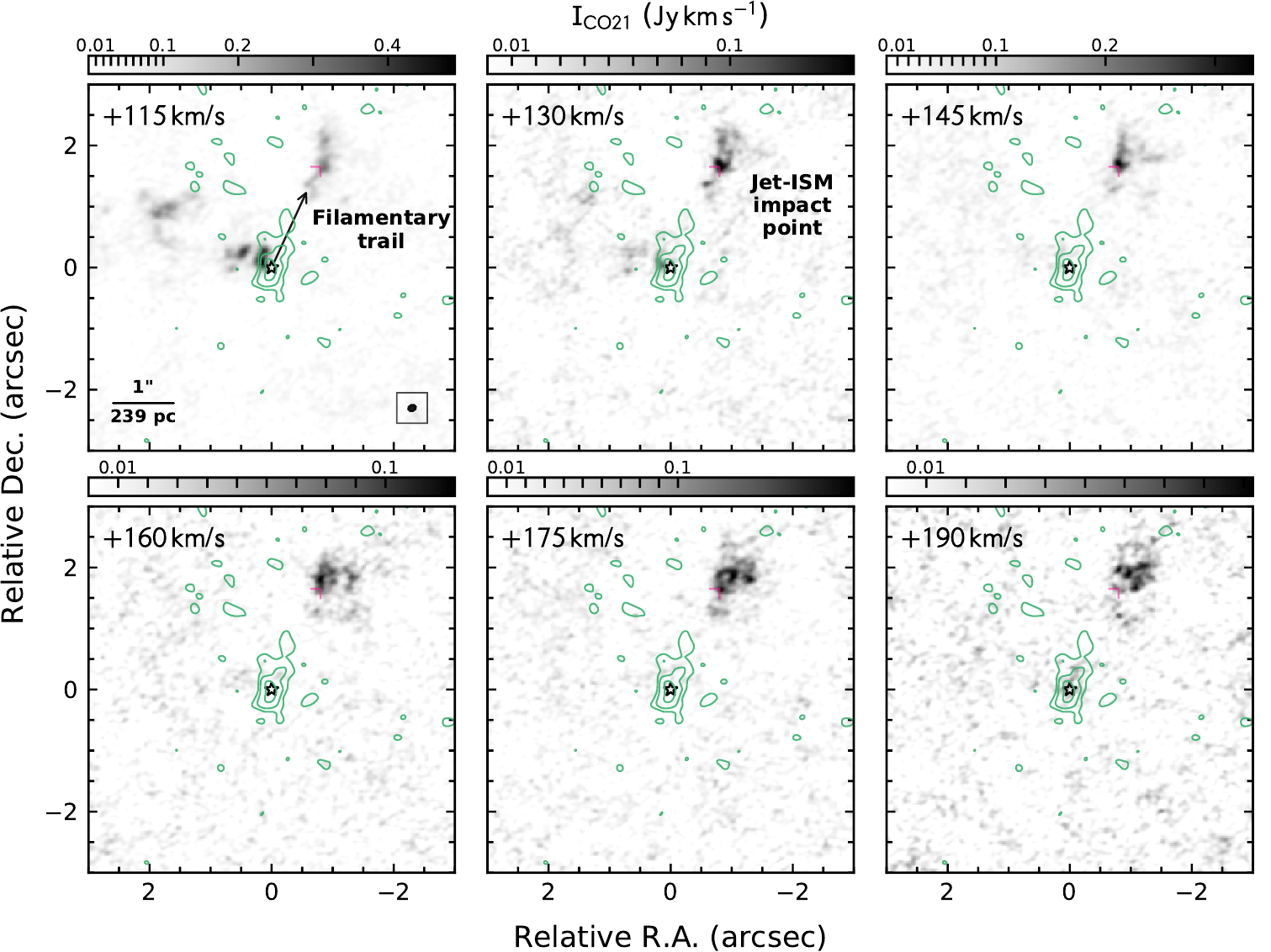}
\caption{\coto\ channel maps from $+115\,\rm{\kms}$ to $+190\,\rm{\kms}$ (step of $15\,\rm{\kms}$; projected velocities) relative to a systemic velocity of $v_{\rm sys} = 3568 \pm 7\, \rm{\kms}$ (this work). A complex filamentary emission on a gradually denser ISM is revealed within the outflowing wind, starting with a diffuse emission trail at $\Delta \alpha = -0\farcs60$, $\Delta \delta = +1\farcs30$ (black arrow; projected distance of $340\, \rm{pc}$ from the AGN) followed by a jet-ISM impact point at $\Delta \alpha = -0\farcs80$, $\Delta \delta = +1\farcs65$ (pink marker; projected $440\, \rm{pc}$). After the bifurcation the wind expands developing a cone-like structure where the highest velocities are found along its central axis. The green-solid contours correspond to the integrated \neii \ emission from VLT/VISIR (starting from $3 \times$\,\textsc{rms} with a spacing of $\times 10^{N/4}$). The black star indicates the position of the AGN in the $1.2\, \rm{mm}$ continuum map. The synthesised beam size is shown in the lower-right corner of the first panel.}\label{fig_chans}
\end{figure*}

\begin{figure}
\centering
\includegraphics[width = \columnwidth]{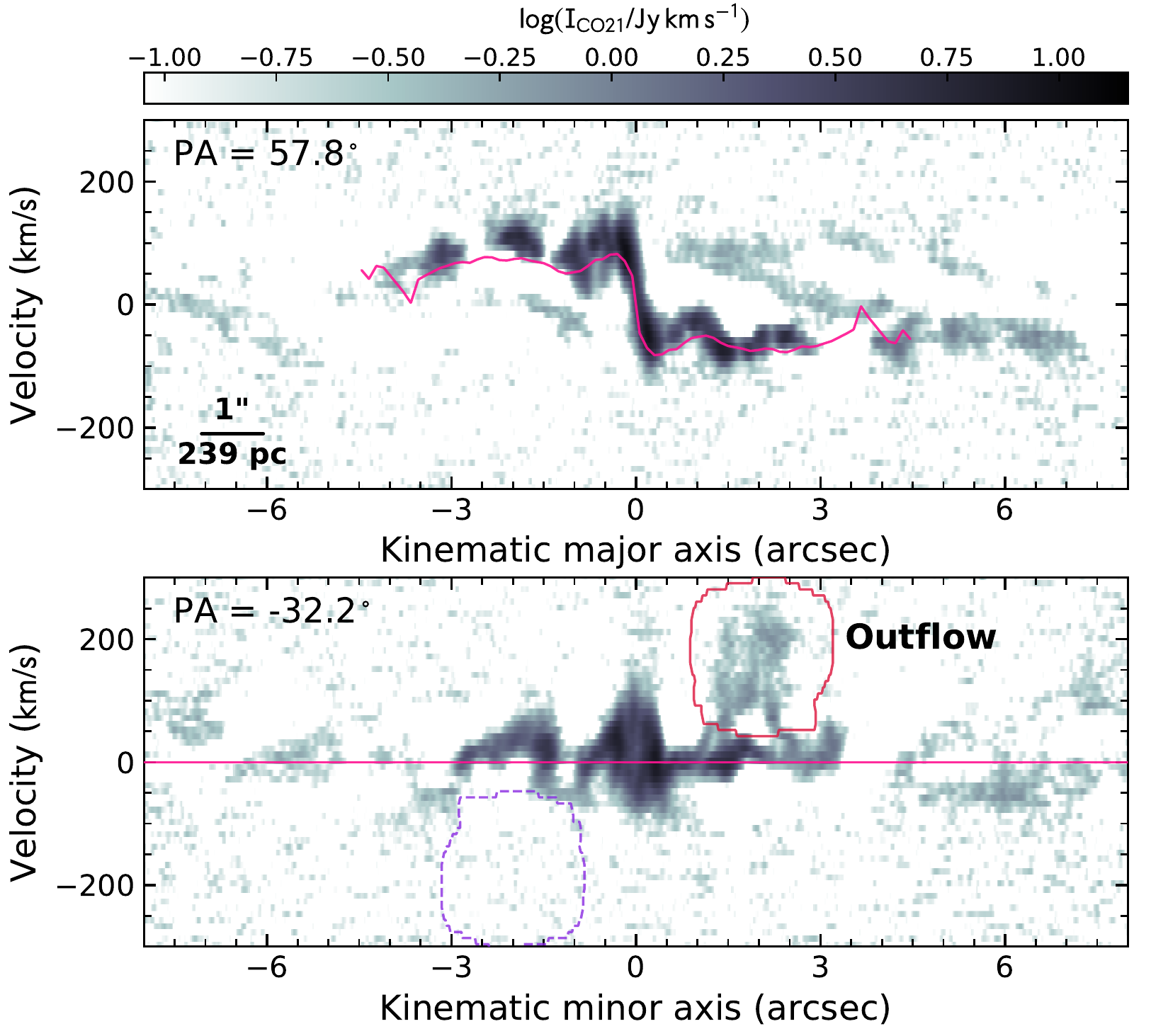}
\caption{\textit{Top:} PV diagram along the kinematic major axis (\textit{PA}\,$= 57\fdg8$). The magenta line indicates the best-fit circular rotation curve obtained with \textsc{Diskfit}. \textit{Bottom:} PV diagram along the kinematic minor axis (\textit{PA}\,$= -32\fdg2$). The magenta horizontal line indicates the best-fit systemic velocity of the galaxy ($3568 \pm 7\, \rm{\kms}$). The solid-red and dashed-purple contours delineate the section of the 3D mask used to integrate the outflow region and its symmetric region across the kinematic minor axis, respectively.}\label{fig_pvmaps}
\end{figure}

\begin{figure*}
\centering
\includegraphics[width = 0.5\textwidth]{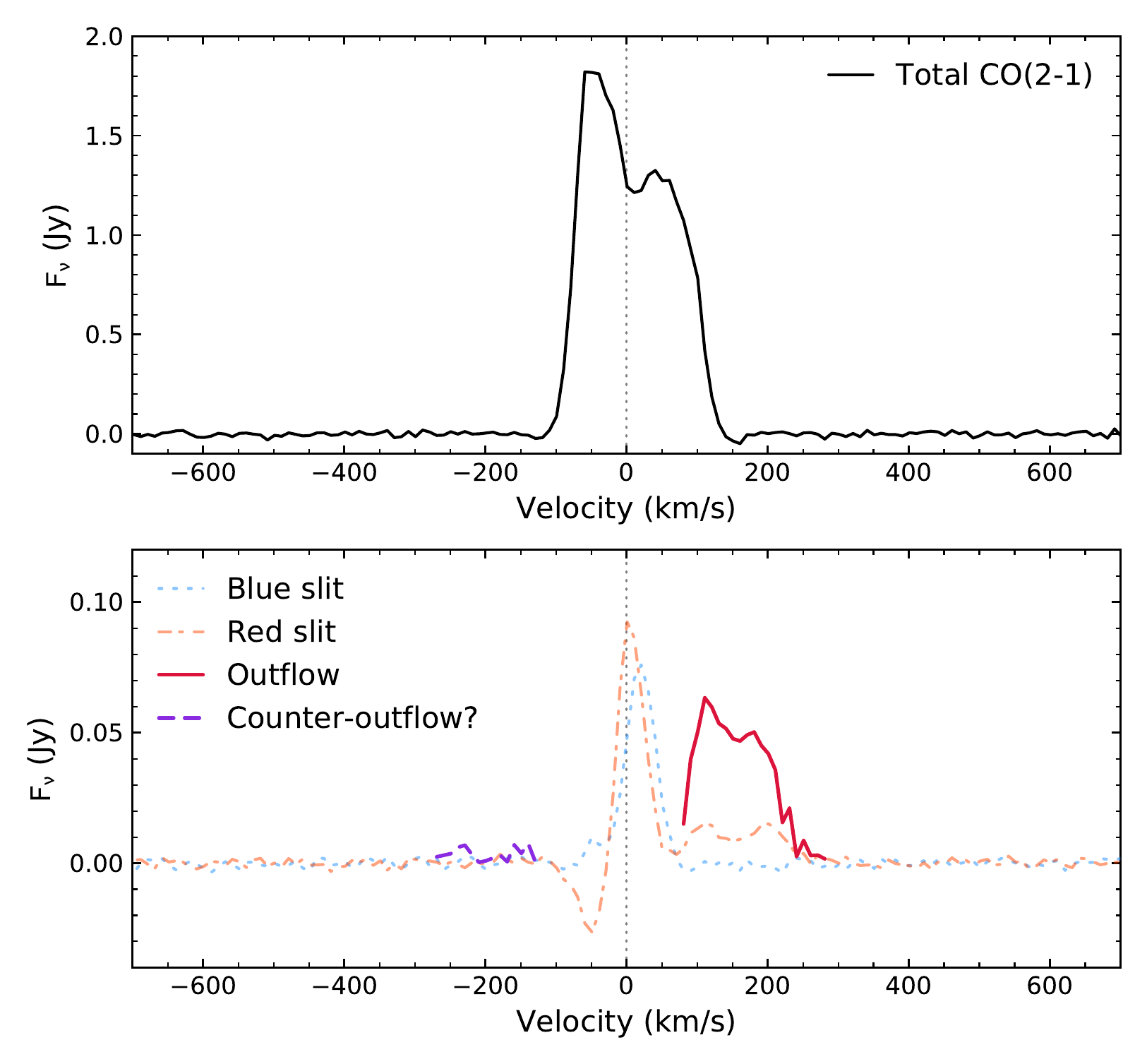}~
\includegraphics[width = 0.5\textwidth]{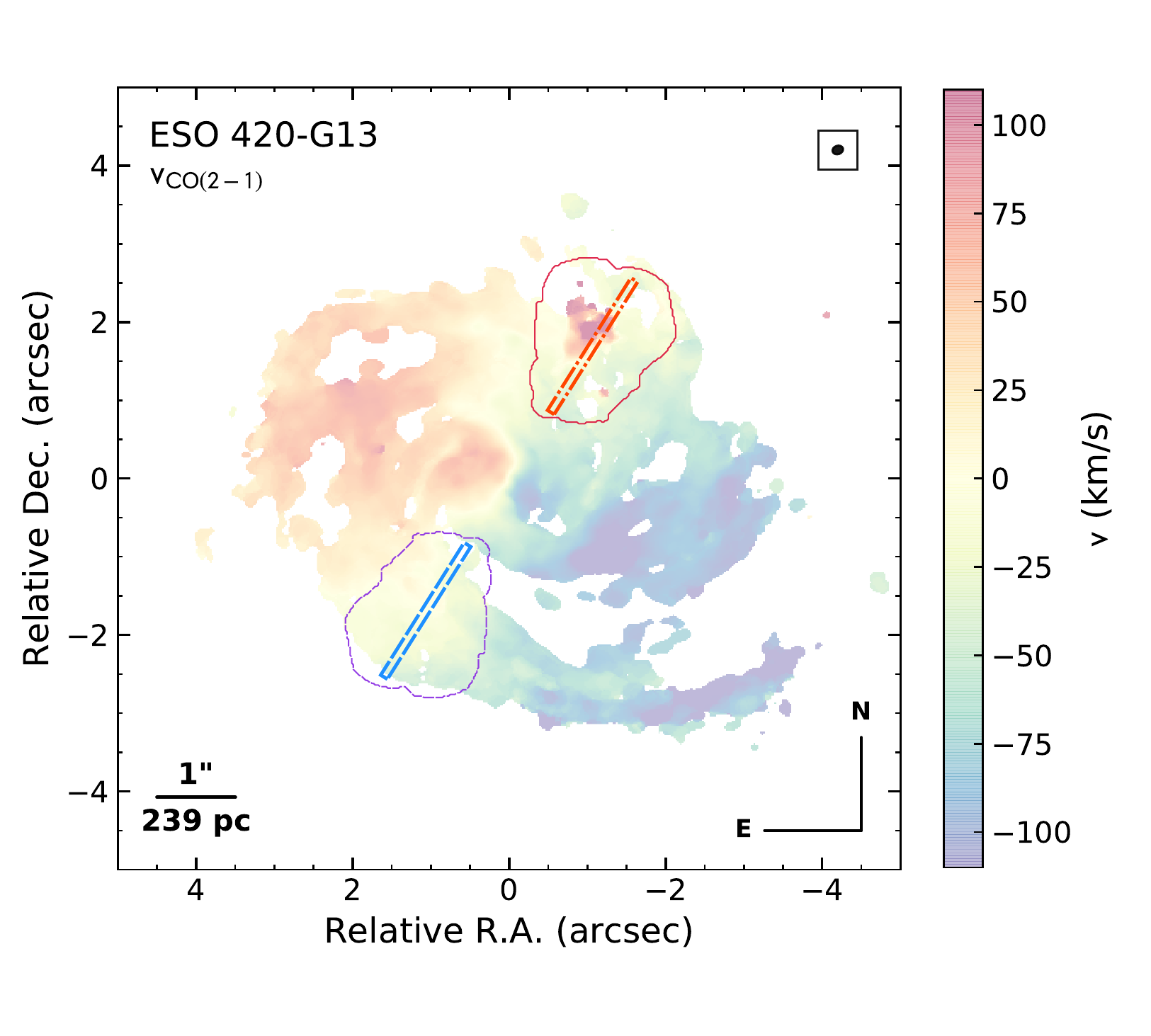}
\caption{\textit{Upper left:} Spectrum of the total integrated CO(2--1) emission for ESO\,420-G13. \textit{Bottom left:} Spectra of the integrated flux for two pseudo-slits extracted along the kinematic minor axis, one intersecting the outflow ($1'' < \Delta x < 3''$, orange dot-dashed line) and one in the opposite direction ($-3'' < \Delta x < -1''$, blue-dotted line). The CO(2--1) flux extracted from a 3D mask applied to the ALMA datacube allows us to isolate the outflow emission (red-solid line) from the rotating gas. A symmetric mask located South-East of the nucleus at blueshifted velocities has also been applied, although no counter-outflow is detected in the integrated spectrum (purple-dashed line). Note that the negative amplitudes at $\sim -70\, \rm{\kms}$ are produced by interferometric artefacts in the image reconstruction process. \textit{Right:} The pseudo-slit positions (blue dashed and orange dot-dashed rectangles) and the projections of the outflow mask (red-solid contour) and the counter-outflow mask (purple-dashed contour) are indicated on the mean velocity map.}\label{fig_spec}
\end{figure*}

\subsection{A molecular gas wind}\label{CO_wind}
A remarkable wind has been detected thanks to its redshifted CO(2--1) emission extended along the kinematic minor axis of ESO\,420-G13, within the inner kpc of the galaxy. It can be clearly distinguished in the mean velocity and sigma moment maps obtained from the spectral cube and shown in the lower panels of Fig.\,\ref{fig_moments}. The wind extends from $\sim 1\farcs5$ to $2\farcs5$ away from the nucleus ($340\, \rm{pc}$ to $600\, \rm{pc}$) with projected velocities in the $+50$--$250\, \rm{\kms}$ range, and presumably indicates the orientation of the central engine axis. Its locus with respect to the systemic velocity ($3568 \pm 7\, \rm{\kms}$, this work) can be seen in the cube slices from $115$ to $190\, \rm{\kms}$ shown in Fig.\,\ref{fig_chans}. The channel map reveals a characteristic funnel shape morphology (e.g. \citeads{2008ApJS..175..423P,2011ApJ...728...29W}), with a narrow filamentary emission starting at $\Delta \alpha = -0\farcs60$, $\Delta \delta = +1\farcs30$ ($340\, \rm{pc}$ projected distance), at $115\,\rm{\kms}$, elongated in the outflow direction, prior to a bifurcation point. The latter is probed by a bright spot in CO(2--1) emission at $130$ and $145\,\rm{\kms}$, located at $\Delta \alpha = -0\farcs80$, $\Delta \delta = +1\farcs65$ relative to the $1.2\, \rm{mm}$ continuum nucleus (pink marker in Fig.\,\ref{fig_chans}; $440\, \rm{pc}$ projected distance). At this forking point, the wind starts spreading further out perpendicularly to the direction of motion, developing a complex structure formed by filaments and several individual clouds that move faster along the cone axis and also with increasing distance from the nucleus. The width of the cone, measured in the direction perpendicular to the kinematic minor axis, ranges from the size of a beam/individual cloud ($\sim 20\, \rm{pc}$) up to $200\, \rm{pc}$ in its widest point. This morphology is also traced by the sigma map in Fig.\,\ref{fig_moments} ($\sigma \sim 50\, \rm{\kms}$).

To determine the position angle (\textit{PA}) of the kinematic major and minor axes, we have modelled the galaxy rotation of the CO(2--1) gas using \textsc{DiskFit}\footnote{\url{https://www.physics.queensu.ca/Astro/people/Kristine_Spekkens/diskfit/}} \citepads{2015arXiv150907120S}. This code applies a $\chi^2$ minimisation to find a global model that fits the circular speed of the disc measured at different radii. In order to avoid over-fitting of non-axisymmetric components, only the rotation was considered here, thus radial flows or non-circular motions were not included in the fit. Furthermore we fixed the axis ratio at $b/a = 0.96$ (from NED\footnote{The NASA/IPAC Extragalactic Database (NED) is operated by the Jet Propulsion Laboratory, California Institute of Technology, under contract with the National Aeronautics and Space Administration.} based on 2MASS images). The best-fit model has a \textit{PA}\,$= 57\fdg8 \pm 0\fdg1$ for the kinematic major axis, and a systemic velocity of $v_{\rm sys} = 3568 \pm 7\, \rm{\kms}$\footnote{Relativistic velocity, defined in the kinematic local standard of rest frame (LSRK).}. The rotation curve obtained is shown in the position-velocity (PV) diagram in Fig.\,\ref{fig_pvmaps} (pink line in top panel). The molecular gas motions within the innermost $r \lesssim 0\farcs4$ ($95\, \rm{pc}$) appear to be dominated by a nuclear ring or disc which coincides with the ring observed in both the dust continuum and the CO(2--1) intensity distribution (see insets in Fig.\,\ref{fig_moments}). Outside of the ring, the velocities drop by $\sim 50\, \rm{\kms}$ from $r = 0\farcs4$ to $1''$ at both sides of the major axis, as shown by the upper panel in Fig.\,\ref{fig_pvmaps}, to increase again up to $80$--$100\, \rm{\kms}$ at $\sim 1\farcs5$ from the nucleus. This is also evident in the mean velocity map in Fig.\,\ref{fig_moments}. The secondary maxima in velocity occur at the intersection of the PV cuts along the major axis with the spiral arms, suggesting that streaming motions associated with the density waves in the galaxy disc alter the local velocity field in these regions.

The spectral profile of the total integrated CO(2--1) emission shown in Fig.\,\ref{fig_spec} (upper panel) is dominated by the galaxy rotation, which has a larger contribution to the intensity in the blue side of the profile. The outflow has a minor contribution to the integrated intensity ($\lesssim 3\%$), while it is easily identified in the slice of the CO(2--1) datacube along the kinematic minor axis (bottom panel in Fig.\,\ref{fig_pvmaps}). 
The wind can be seen in the spectrum of a short slice extracted along the kinematic minor axis ($1'' < \Delta x < 3''$; orange dot-dashed line in Fig.\,\ref{fig_spec}, bottom-left panel), however the slit intersects only a small section of the outflow cone (see slit positions in the right panel of Fig.\,\ref{fig_spec}). Thus, the amplitude of the extracted spectrum is comparable to the interferometric artefacts in the map (see the negative residual at blueshifted velocities). An optimal extraction of the outflow spectrum can be performed using a 3D mask in the ALMA datacube, to select the outflow region in the spatial dimensions and isolate the wind from the rotating gas in the velocity dimension. The red contours in Figs\,\ref{fig_pvmaps} and \ref{fig_spec} represent the different projections of the 3D mask on the PV diagram and the image plane, respectively, while the integrated spectrum is shown in Fig.\,\ref{fig_spec} (red-solid line). No counter outflow is detected at the blueshifted side of the kinematic minor axis ($-3'' < \Delta x < -1''$; blue-dashed line in Fig.\,\ref{fig_spec}) nor in the integrated spectrum of the opposite symmetric 3D mask (purple-dashed line). See Section\,\ref{discuss} for a further discussion on this issue.

The integrated wind emission ($7.0 \pm 0.6\, \rm{Jy\,\kms}$) translates into a molecular gas mass of $(8.3 \pm 0.7) \times 10^6\, \rm{\msun}$, derived from the \citetads{1997ApJ...478..144S} formula and $\alpha = 0.8\, \rm{M_\odot\,(K\,\kms\,pc^2)^{-1}}$, typical of LIRG and ULIRG galaxies \citepads{2013ARA&A..51..207B}. As no information on the excitation and optical depth of the gas in the wind is available, it is unclear whether other $\alpha$ values, such as those found for IC\,5063 \citepads{2016A&A...595L...7D}, could be appropriate. If instead a Milky Way value of $\alpha = 4.3\, \rm{M_\odot\,(K\,\kms\,pc^2)^{-1}}$ is assumed \citepads{2013ARA&A..51..207B}, the derived mass would be significantly larger, of $(4.5 \pm 0.4) \times 10^7\, \rm{\msun}$. For the case of optically thin emission, the wind mass would drop by a factor of $\sim 3$ (e.g. \citeads{2013A&A...558A.124C,2016A&A...595L...7D}).

\subsection{Extended ionised gas}\label{ion_wind}
The \neii \ emission-line contours above $3 \times$\,\textsc{rms} have been over-plotted on the moment maps in Fig.\,\ref{fig_moments} and the channel maps in Fig.\,\ref{fig_chans}. The ionised gas distribution shows a central peak --\,possibly associated with the AGN\,-- plus an elongated plume with an extension of about $1''$ ($240\, \rm{pc}$) $3 \times$\,\textsc{rms} along the North-West direction, coincident with the kinematic minor axis. The orientation of the ionised gas emission is surprisingly well aligned with that of the CO(2--1) outflow, as best seen in e.g. the $115\, \rm{\kms}$ velocity slice (upper-left panel in Fig.\,\ref{fig_chans}). The \textit{Spitzer}/IRS spectrum obtained by \citetads{2015ApJS..218...21L} suggests that the continuum filter is not affected by contamination from PAH or blueshifted \neii \ emission (see Fig.\,\ref{fig_ne2}).


At scales larger than the $3 \times 3\, \rm{kpc^2}$ covered by the ALMA observations, ESO\,420-G13 presents an extended tail of ionised gas towards the North-East and possibly the South-West, produced by the AGN and the star formation \citepads{2017ApJS..232...11T}. The far-IR lines observed by \textit{Herschel}/PACS \citepads{2016ApJS..226...19F} show a brighter distribution along the kinematic minor axis (e.g. see the [\ion{O}{i}]$_{\rm 63 \mu m}$, the [\ion{O}{iii}]$_{\rm 88 \mu m}$, and the [\ion{C}{ii}]$_{\rm 158 \mu m}$ maps in Fig.\,\ref{fig_o3_c2}). The lower angular resolution in the far-IR ($9\farcs4 \times 9\farcs4$ per spaxel vs. $1\farcs7$ seeing in the optical) does not explain this difference. No dust lanes that could explain this displacement are detected in the IRAC (Fig.\,\ref{fig_zoom}) or the 2MASS archival images. A more efficient cooling of the diffuse rarefied gas through the far-IR lines \citepads{O&F06} needs to be examined using observations at higher angular resolution.




\section{Discussion}\label{discuss}

\subsection{A wind revealing a jet}\label{windjet}
The results shown in Section\,\ref{results} point out to the detection of a jet-related wind in ESO\,420-G13, i.e., of a collimated outflow oriented along the minor axis of the galaxy and powered by the mechanical energy input from the AGN. In this picture, we interpret the extended \neii\ emission as a collimated ionised wind associated with hot gas in the jet cavity. The structure of the molecular gas outflow further out resembles closely that of numerical simulations of jet-cloud interactions \citepads{2011ApJ...728...29W,2016AN....337..167W}. The jet percolates the porous ISM propagating through a tenuous medium until it collides with a denser molecular core and splits, leading to the apparent bifurcation. Fig.\,\ref{fig_chans} shows a collimated, filamentary outflow starting at a projected distance of $340\, \rm{pc}$ from the nucleus and at relatively low velocities ($115\, \rm{\kms}$). At higher velocities, the outflow obtains a conical shape, with a starting point that is most likely associated with a dense jet-ISM impact point at $440\, \rm{pc}$. The exterior of the dispersed gas cone has lower projected velocities than the interior of the cone, which is sensible if, e.g., part of the jet remains in its initial trajectory. The dispersed gas also moves faster ($160$--$290\, \rm{\kms}$; Fig.\,\ref{fig_chans}) with increasing distance from the impact point, which is sensible if energy is mainly transported near the jet front. Inversely, the narrow side of the cone shows relatively high velocity dispersion ($\gtrsim 50\, \rm{\kms}$) when compared to the gas rotation ($\sim 20\, \rm{\kms}$), which could be explained by the effect of higher turbulence  \citepads{2012ApJ...757..136W} or higher ram pressure (in the case of a two-phase medium; \citeads{2010MNRAS.401....7H}) in the clouds that first experienced the jet impact. Overall, overpressured jets can inflate a bubble within the ISM, bifurcate and cause conical winds and disperse the molecular gas into several clumps and filaments \citepads{2012ApJ...757..136W}. This picture is similar to that seen in the galaxy IC\,5063 \citepads{2016A&A...595L...7D}.

The cases of NGC\,1377 and ESO\,420-G13 suggest that unresolved jets might still have an important role in the feeding process even for non powerful radio galaxies. In NGC\,1377 \citepads{2016A&A...590A..73A}, no radio emission was known at the time of the molecular wind detection. In ESO\,420-G13, faint radio emission was seen, but its origin could not be linked to a jet. The molecular wind suggests the presence of a jet, which was most likely unresolved in previous radio observations ($< 17''$; \citeads{1996ApJS..103...81C,1998AJ....115.1693C}). Overall, elusive jets like those seen in NGC\,1377 or ESO\,420-G13 can be detected through their interaction with the ISM. While in NGC\,1377 a continuous outflow can be traced from the nucleus up to a distance of $150\, \rm{pc}$ at both sides along the jet axis, in ESO\,420-G13 the molecular gas wind starts at $340\, \rm{pc}$ away from the nucleus, which could further strengthen the argument of the presence of a jet. At shorter radii the molecular gas shows regular rotation, while the morphology of the \neii \ emission suggests that and ionised wind might be present in the innermost few hundred parsecs. Future integral field spectroscopic observations at subarcsec resolution would be required to confirm the nature of the ionised wind. No counter-outflow is detected in the opposite direction of the redshifted molecular gas outflow. In the following Subsection, we propose geometries that could explain the jet-cloud interactions.

\subsection{Jet-cloud configuration}\label{scenarios}
Two main scenarios are explored to explain how the observed outflow could be driven by jet-cloud interactions. First, we consider a jet close to the galaxy minor axis, colliding with a cloud located far from the galaxy plane. The non-detection of a blueshifted counterpart is reasonable, since the chance to find a molecular cloud at a relatively high altitude is low. The existence of such a cloud could even require out-of-equilibrium dynamical conditions, caused by e.g. by a previous feedback episode or by a minor merger in the past. A previous outflow event could have been launched by SNe, or by the AGN radiation, or by former jet activity, as in e.g. the case of Centaurus\,A \citepads{2016A&A...595A..65S}. In this scenario, the molecular gas would have reformed as the relic outflow cooled down \citepads{2018MNRAS.474.3673R}. Thus, the current jet event could push further away the same material that was ejected in the past. Alternatively, the minor merger scenario could be supported by the high IR luminosity in \mbox{ESO\,420-G13} ($\sim 10^{11}\, \rm{L_\odot}$) and the post-starburst features reported by \citetads{2017ApJS..232...11T} in the optical spectrum. This could also explain the existence of molecular gas clouds at a high altitude above the disc and the non-symmetric distribution of clouds in the opposite side. However, no stellar emission in the IRAC $3.6\, \rm{\micron}$ associated with the minor companion has been detected (see left panel in Fig.\,\ref{fig_zoom}). With the current dataset, we deem the past feedback event more likely than the past minor merger event.

In the second scenario, the jet is pushing molecular gas that belongs to the (thick) galaxy disc, at an intrinsically low altitude. For a jet moving nearly parallel to the disk due to precession, as in the case of IC\,5063, the jet does not really shatter until it hits a dense molecular core at $340\, \rm{pc}$. At shorter radii the jet percolates the ISM through the inner circumnuclear ring, leading to the \neii \ emission in the low-density cavities. However, the outflow asymmetry is hard to justify. Otherwise, one can consider a jet emerging perpendicular to the galaxy disc, which then bents or precesses at a certain height and gets redirected towards the disc. This would explain why the innermost gas is unaffected. The \neii \ emission in this case would be also associated with the low-density ionised medium cone carved by the jet in the ISM. The one-sided wind would require an asymmetric jet bending, which is often observed in radio galaxies. However, the jet bending region is not identified in the CO(2--1) or the \neii \ line maps. If the jet bent was caused by precession, the symmetric counter-jet would have likely generated a blueshifted outflow, since the disc is rich in molecular gas near the expected impact point. If the jet bent was linked to a collision with a molecular cloud, then this cloud is small or destroyed and, thus, non-detectable.

\subsection{Energy and momentum balance}
Another argument favouring the jet scenario comes from the comparison of the wind kinetic luminosity and the radio power of ESO\,420-G13. The wind kinetic luminosity ($L_{\rm kin}$) was computed from the product $\frac{1}{2} M V^{3} d^{-1}$, where $M$ is the mass of the gas in the wind, $V$ is the wind speed measured as the average speed of the outflow spectrum in Fig.\,\ref{fig_spec} ($160\, \rm{\kms}$, projected velocity), and $d$ is the distance from the location of the driving mechanism. For a radio jet, this is equivalent to the jet-cloud impact point. In our computations, we always assume that the wind started away from the nucleus, not that it got transported from the nucleus to $\sim 340\, \rm{pc}$ away. Considering that the jet-ISM interaction occurs near the forking point of the outflowing CO(2--1) emission (Fig.\,\ref{fig_chans}), then the average projected distance covered by the wind is $d \sim 100\, \rm{pc}$ ($0\farcs4$). For an outflow mass of $(8.3 \pm 0.7) \times 10^6\, \rm{\msun}$ (see Section\,\ref{CO_wind}), we derive a mass outflow rate of $\dot{M} \sim M V/d = 14 \pm 1\, \rm{M_\odot\,yr^{-1}}$, a kinetic luminosity of $L_{\rm kin} = 1.1 \times 10^{41}\, \rm{erg\,s^{-1}}$, and a momentum rate of the accelerated gas of $\dot{M} V = 1.4 \times 10^{34}\, \rm{erg\,cm^{-1}}$. The error in the mass outflow rate corresponds only to the propagated uncertainty of the CO(2--1) intensity measurement, while a larger uncertainty is expected from the assumed geometry and the adopted $\alpha$ value. Furthermore, these values should be considered as lower limits due to the projection effect. If an inclination angle of $i \sim 20\degr$--$30\degr$ is assumed, then the deprojected velocity would be of about $320$--$470\, \rm{\kms}$, and the corresponding kinetic luminosity would be $\sim 4$--$9 \times 10^{41}\, \rm{erg\,s^{-1}}$. Note that we are not including the velocity dispersion to define the average velocity of the wind, as has been done in e.g. \citetads{2019MNRAS.483.4586F}. Although this approach may be appropriate for a spherical wind, given the high collimation in our case we consider that including the turbulence of the wind would overestimate its average velocity.

\begin{figure}
\centering
\includegraphics[width = \columnwidth]{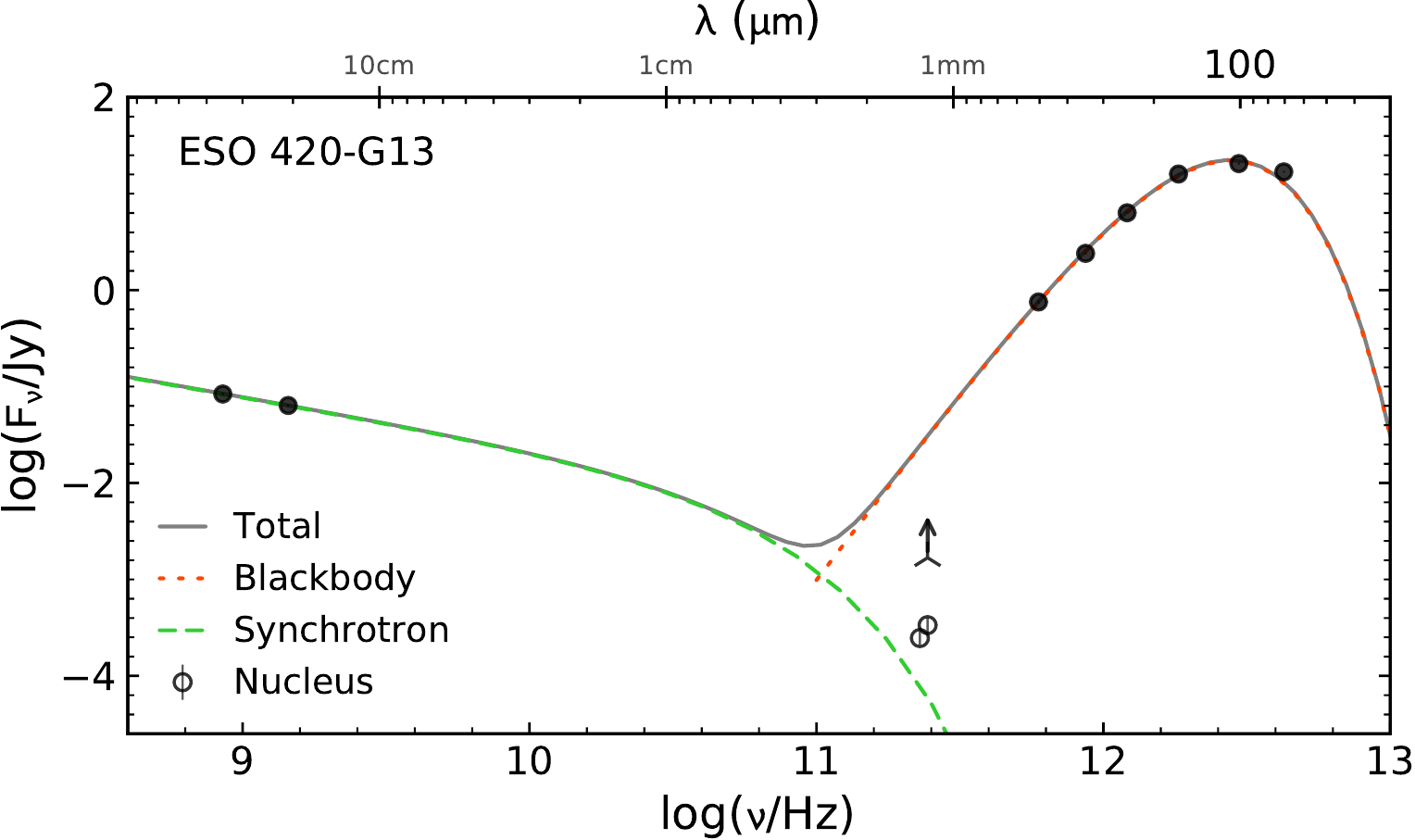}
\caption{Flux distribution of the radio-to-IR continuum emission for ESO\,420-G13 compiled from the literature (black circles). Our ALMA continuum measurements at $227\, \rm{GHz}$ and $241\, \rm{GHz}$ for the nuclear point-like source ($< 0\farcs1$) are represented by open circles, while the total integrated flux within the inner $27''$ at $241\, \rm{GHz}$ is represented as a lower limit, due to possible spatial filtering of flux extended over spatial scales larger than $0\farcs8$ in the interferometric observations (see Table\,\ref{tab_sed}). The model including synchrotron continuum (green-dashed line) and blackbody emission (orange-dotted line) has been fitted to the flux distribution (grey-solid line). Note that flux errors do not appear in the figure due to the smaller size when compared to the plot symbols.}\label{fig_sed}
\end{figure}

We first compare these quantities with the kinetic and radiative power of the active nucleus in ESO\,420-G13. In order to compute the radio power we used two different methods that provide lower and upper limits. The lower limit is derived by fitting the radio-to-IR spectral energy distribution (SED), compiled from NED in the $843\, \rm{MHz}$ to $70\, \rm{\micron}$ range (Fig.\,\ref{fig_sed}; Table\,\ref{tab_sed}). The dust component was fitted using a modified black body \citepads{2012MNRAS.425.3094C}. Assuming that the synchrotron radiation from the nucleus dominates at low frequencies, this component was modelled using a broken power law with an exponential cut off \citep{R&L04}, which is in agreement with the ALMA fluxes for the radio core (open circles in Fig.\,\ref{fig_sed}, not included in the fit). Using a standard $\chi^2$ minimisation method, a minimum radio power of $10^{39}\, \rm{erg\,s^{-1}}$ was found. The jet power, as computed from its radio emission, could be missing some of the energy deposited in a hot bubble that does pdV work. An estimate of the total mechanical power is obtained from the calibration by \citetads{2014ARA&A..52..589H} for a sample of $\sim 40$ radio galaxies, comparing the derived values of the work required to inflate the X-ray cavities observed in these galaxies with the monochromatic $1.4\, \rm{GHz}$ continuum flux of the nucleus (data from \citeads{2010ApJ...720.1066C}, \citeads{2008ApJ...686..859B}, and \citeads{2006ApJ...652..216R}). Following \citetads{2014ARA&A..52..589H} we use a normalisation factor of $f_{\rm cav} = 4$, and a $1.4\, \rm{GHz}$ flux of $65\, \rm{mJy}$ (\citeads{1998AJ....115.1693C}; Table\,\ref{tab_sed}) to estimate a total mechanical power of $\sim 4 \times 10^{42}\, \rm{erg\,s^{-1}}$ in ESO\,420-G13. Therefore, the radio power differs from the wind kinetic luminosity by up to an order of magnitude, despite the wide range in the estimated jet power.

For a black hole mass of $M_{\rm BH} \sim 4 \times 10^8\, \rm{M_\odot}$ (see Section\,\ref{intro}), and an AGN luminosity of $L_{\rm AGN} = 0.5$--$1 \times 10^{44}\, \rm{erg\,s^{-1}}$ --\,that is $10$--$20\%$ of the total IR luminosity based on the flux ratios of the mid-IR fine-structure lines observed by \textit{Spitzer}/IRS (\nev/\neii\ $= 0.10 \pm 0.01$ and \oiv/\neii \ $= 0.45 \pm 0.03$; \citeads{2016ApJS..226...19F}), the corresponding Eddington ratio is $\log(L_{\rm AGN}/L_{\rm edd}) \lesssim -2.7$. This suggests that the active nucleus is likely in the radio or kinetic mode, thus favouring the jet scenario. Notably, ESO\,420-G13 is not classified as radio-loud according to classical IR-to-radio flux ratios integrated over the whole galaxy ($q_{\rm IR} = 2.7$, radio-loud have $q_{\rm IR} \lesssim 1.8$; \citeads{2010MNRAS.402..245I}). Just considering the AGN contribution would lower $q_{\rm IR}$ to values in the $1.7$--$2.0$ range, thus the nucleus would be very close to the radio-loud domain if the radio emission is dominated by the AGN. Higher angular resolution observations ($< 1''$) at radio wavelengths are required to probe the jet morphology and constrain its power estimate.

\subsection{Other wind drivers}
How do we know that the wind is not triggered by AGN radiation? The AGN emission accounts only for $\sim 10$--$20\%$ of the galaxy's bolometric luminosity, that is $L_{\rm AGN} = 0.5$--$1 \times 10^{44}\, \rm{erg\,s^{-1}}$. This is more than three orders of magnitude higher when compared to the X-ray luminosity, which could be instead dominated by star formation in the host galaxy (see Section\,\ref{intro}). In case the wind was driven by radiation pressure from the AGN, its distance from the generating mechanism would be higher than before, i.e. $\sim 600\, \rm{pc}$ ($2\farcs5$). Thus, the wind mass flow rate and momentum rate would be $2.3\, \rm{\msun\,yr^{-1}}$ and $2.3 \times 10^{33}\, \rm{erg\,cm^{-1}}$, respectively. This momentum rate does not exclude the AGN as a potential driving mechanism, considering the estimated radiation pressure of $L_{\rm AGN}/c = 1.7$--$3.7 \times 10^{33}\, \rm{erg\,cm^{-1}}$ and the momentum boosting that can be achieved during some phases of energy-driven expansion \citepads{2014A&A...562A..21C}. Still, in this scenario it would be hard to explain why the molecular gas rotating close to the nucleus is not blown away.

Regarding the radiation pressure from stars, no clear sign of a local starburst --\,such as excess emission in the dust continuum\,-- is detected near the outflow region. No significant excess of hot dust can be claimed either in the $12.8\, \rm{\micron}$ continuum image. Rather weak emission is seen in the $1.2\, \rm{mm}$ continuum map (see upper-left panel in Fig.\,\ref{fig_moments}), originating from three discrete knots in or near the area that the CO wind occupies. These regions all have comparable fluxes, in the $0.11$--$0.15\, \rm{mJy}$ range, contributing up to a total of $0.48\, \pm 0.08\, \rm{mJy}$. Even if all of the pertinent emission was originating from the cold dust instead of the jet-related synchrotron, none of these regions could produce more than $2.5\%$ of the total cold dust emission, which is equal to $6.1 \pm 0.3\, \rm{mJy}$. The corresponding infrared luminosity fraction is $2 \times 10^9\, \rm{\lsun}$ (as \lir \ for the entire galaxy is $8 \times 10^{10}\, \rm{\lsun}$). From \citetads{1998ApJ...498..541K}, this translates into a local star formation rate (SFR) of $0.3\, \rm{\msun\,yr^{-1}}$. Given that $\sim 100\, \rm{\msun}$ are needed for one supernova (SN) event to take place, the probability is very low for a SN to be responsible for this wind. Alternatively, a population of (young) stars cannot drive the wind either. The area comprising the CO wind occupies $4\%$ of the overall galaxy in the near infrared (JHKs 2MASS bands). The overall galaxy emission, which is comparable in the IR and in the optical bands and includes the AGN contribution, is $(5.8 \pm 1.3) \times 10^{44}\, \rm{erg\,s^{-1}}$. The force exerted to the gas due to stellar radiation pressure, $L_{\rm stars}/c$, would then be as high as $7.7 \times 10^{32}\, \rm{erg\,cm^{-1}}$, significantly lower that the momentum rate of the gas ($1.4 \times 10^{34}\, \rm{erg\,cm^{-1}}$). Therefore, neither SNe nor stellar radiation can locally drive the wind.
\begin{table}
\caption{Kinetic luminosity carried away by the molecular gas wind in ESO\,420-G13, compared to different possible launching mechanisms, i.e. AGN radiation, jet power, and star formation. The latter correspond to the knots identified in the ALMA $1.2\, \rm{mm}$ continuum located close to the outflow wind.}\label{tab_en}
\centering
\begin{tabular}{lc}
  Component  & Luminosity \\
             & ($\rm{erg\,s^{-1}}$) \\
  \hline \\[-0.3cm]
  Outflow    & $1.1 \times 10^{41}$ \\
  AGN        & $0.5$--$1 \times 10^{44}$ \\
  Jet        & $0.001$--$4 \times 10^{42}$ \\
  Starburst  & $7.7 \times 10^{42}$ \\
  \hline
\end{tabular}
\end{table}

\subsection{Further evolution of ESO\,420-G13}\label{evol}
For a mass outflow rate of $14\, \rm{M_\odot\,yr^{-1}}$ and a total molecular gas mass of $M_{\rm tot} = (3.07 \pm 0.02) \times 10^8\, \rm{M_\odot}$ (see Section\,\ref{results}), the depletion time would be of $\eta = M_{\rm tot}/\dot{M} = 23\, \rm{Myr}$ at the current outflow rate. However the outflowing gas would not likely escape the galaxy, falling again to the central part. Therefore, the result of this interaction points to a delay of the star formation instead of a quenching, while most of the feedback effect is expected in the central few hundred parsecs of the galaxy. This is in line with the results from \citetads{2019MNRAS.483.4586F}. More violent AGN activity in the past would be required in order to explain the current evolutionary stage of ESO\,420-G13.

\section{Summary}\label{sum}

In this work we present a collimated molecular gas outflow detected in the CO(2--1) transition using ALMA interferometric observations of the central $3 \times 3\, \rm{kpc^2}$ in the Seyfert 2 galaxy ESO\,420-G13. The molecular gas outflow has a conical shape with $\sim 20\, \rm{pc}$ width at the closest point to the nucleus up to $200\, \rm{pc}$ width at the farthest point. It carries a molecular gas mass of $\sim 8 \times 10^6\, \rm{M_\odot\, yr^{-1}}$ at an average projected velocity of $160\, \rm{\kms}$, which translates into an outflow rate of $\sim 14\, \rm{M_\odot\,yr^{-1}}$. Based on the outflow structure and the energy and momentum balance, we conclude that this is a jet-driven wind powered by mechanical energy input from the AGN, becoming the second case after NGC\,1377 in which the presence of a previously unknown jet is revealed through its interaction with the ISM. However, in ESO\,420-G13 the molecular gas wind is detected very far from the AGN, at $340\, \rm{pc}$ (projected), traced only by \neii \ ionised gas emission at closer distances down to the nucleus. Two possible scenarios are proposed to explain the origin of such an outflow: \textit{i)} an outer molecular cloud originated by previous jet activity or a minor merger event, then impacted by the current jet, or \textit{ii)} a molecular cloud in the galaxy disc impacted by a jet, either propagating through the porous ISM or after a bent or a precession. The outflowing molecular gas detected will likely fall back into the galaxy, delaying instead of quenching the star forming process in the centre.

The case of ESO\,420-G13 proves that moderate-luminosity jets in non-powerful radio galaxies can still play a main role driving molecular gas outflows in these objects. Deep and high-angular resolution data for CO lines, radio wavelengths, and ionised gas are required to reveal the presence of these jets. In this context, the TWIST survey will provide a census of similar jet-driven outflows in nearby galaxies, allowing us to move from individual galaxy studies to a robust statistical analysis of these phenomena.

\begin{acknowledgements}
The authors acknowledge the referee for his/her useful comments that helped to improve the manuscript. JAFO acknowledges financial support by the Agenzia Spaziale Italiana (ASI) under the research contract 2018-31-HH.0. JAFO and KMD acknowledge financial support by the Hellenic Foundation for Research and Innovation (HFRI), under the first call for the creation of research groups by postdoctoral researchers that was launched by the General Secretariat For Research and Technology (project number 1882). MPS acknowledges support from the Comunidad de Madrid, Spain, through Atracci\'on de Talento Investigador Grant 2018-T1/TIC-11035 and STFC through grants ST/N000919/1 and ST/N002717/1. CR acknowledges support from the CONICYT+PAI, Convocatoria Nacional subvenci\'on a instalaci\'on en la academia, convocatoria a\~no 2017 PAI77170080. This paper makes use of the following ALMA data: ADS/JAO.ALMA\#2017.1.00236.S. ALMA is a partnership of ESO (representing its member states), NSF (USA) and NINS (Japan), together with NRC (Canada), MOST and ASIAA (Taiwan), and KASI (Republic of Korea), in cooperation with the Republic of Chile. The Joint ALMA Observatory is operated by ESO, AUI/NRAO and NAOJ. The National Radio Astronomy Observatory is a facility of the National Science Foundation operated under cooperative agreement by Associated Universities, Inc.
\end{acknowledgements}

\bibliographystyle{aa}
\bibliography{twist_eso420-g13}

\onecolumn
\begin{appendix}

\begin{table}
\section{Radio to far-IR continuum fluxes}

\caption{Far-IR to radio continuum fluxes for the nucleus of ESO\,420-G13. Flux measurements other than ALMA have been compiled from the literature and the NED database. The columns in the table correspond to the facility used for the continuum measurement, the observed frequency, the continuum flux density, the angular resolution of the observations, and the corresponding reference for the measurements. Note that the largest angular scale in the ALMA maps is $0\farcs8$, thus the measured flux should be considered as a lower limit to the integrated flux within the FOV ($27''$).\vspace{0.2cm}}\label{tab_sed}
\centering
\begin{tabular}{lcccc}
Telescope/instrument & Frequency ($\rm{Hz}$) & Flux (Jy) & Aperture & Ref. \\
\hline \\[-0.3cm]
  PACS $70\, \rm{\micron}$   & $4.2029 \times 10^{12}$ & $17.1 \pm 0.3$ & $6\farcs3 \times 5\farcs5$ & \tablefootmark{a} \\
  PACS $100\, \rm{\micron}$  & $2.9294 \times 10^{12}$ & $20.8 \pm 0.5$ & $7\farcs4 \times 6\farcs8$ & \tablefootmark{a} \\
  PACS $160\, \rm{\micron}$  & $1.8052 \times 10^{12}$ & $16.2 \pm 0.3$ & $12\farcs3 \times 10\farcs5$ & \tablefootmark{a} \\
  SPIRE $250\, \rm{\micron}$ & $1.19553 \times 10^{12}$ & $6.43 \pm 0.07$ & $18\farcs5 \times 17\farcs5$ & \tablefootmark{b} \\
  SPIRE $350\, \rm{\micron}$ & $8.56739 \times 10^{11}$ & $2.44 \pm 0.05$ & $25\farcs3 \times 23\farcs7$ & \tablefootmark{b} \\
  SPIRE $500\, \rm{\micron}$ & $5.87111 \times 10^{11}$ & $0.76 \pm 0.03$ & $37\farcs0 \times 34\farcs1$ & \tablefootmark{b} \\
  ALMA $241\, \rm{GHz}$ & $2.41108 \times 10^{11}$ & $ > 1.7 \times 10^{-3}$ & $\sim 27''$ & \tablefootmark{c} \\
  ALMA $241\, \rm{GHz}$ & $2.41108 \times 10^{11}$ & $(340 \pm 60) \times 10^{-6}$ & $0\farcs08 \times 0\farcs10$ & \tablefootmark{c} \\
  ALMA $227\, \rm{GHz}$ & $2.26605 \times 10^{11}$ & $ > 1.3 \times 10^{-3}$ & $\sim 27''$ & \tablefootmark{c} \\
  ALMA $227\, \rm{GHz}$ & $2.26605 \times 10^{11}$ & $(250 \pm 40) \times 10^{-6}$ & $0\farcs09 \times 0\farcs11$ & \tablefootmark{c} \\
  VLA $1.4\, \rm{GHz}$ & $1.425 \times 10^9$ & $(65 \pm 2) \times 10^{-3}$ & $16\farcs6 \times 17\farcs5$ & \tablefootmark{d} \\
  SUMSS $843 \rm{MHz}$ & $8.43 \times 10^8$ & $(85 \pm 3) \times 10^{-3}$ & -- & \tablefootmark{e} \\
  
  \hline
\end{tabular}
\tablefoot{
\tablefoottext{a}{\textit{Herschel}/PACS Point Source Catalog.}
\tablefoottext{b}{\textit{Herschel}/SPIRE Point Source Catalog.}
\tablefoottext{c}{This Work.}
\tablefoottext{d}{\citetads{1998AJ....115.1693C}.}
\tablefoottext{e}{NED database.}
}
\end{table}



\begin{figure}
\section{Transmission of VISIR narrow-band filters}
\centering
\includegraphics[width = 0.6\textwidth]{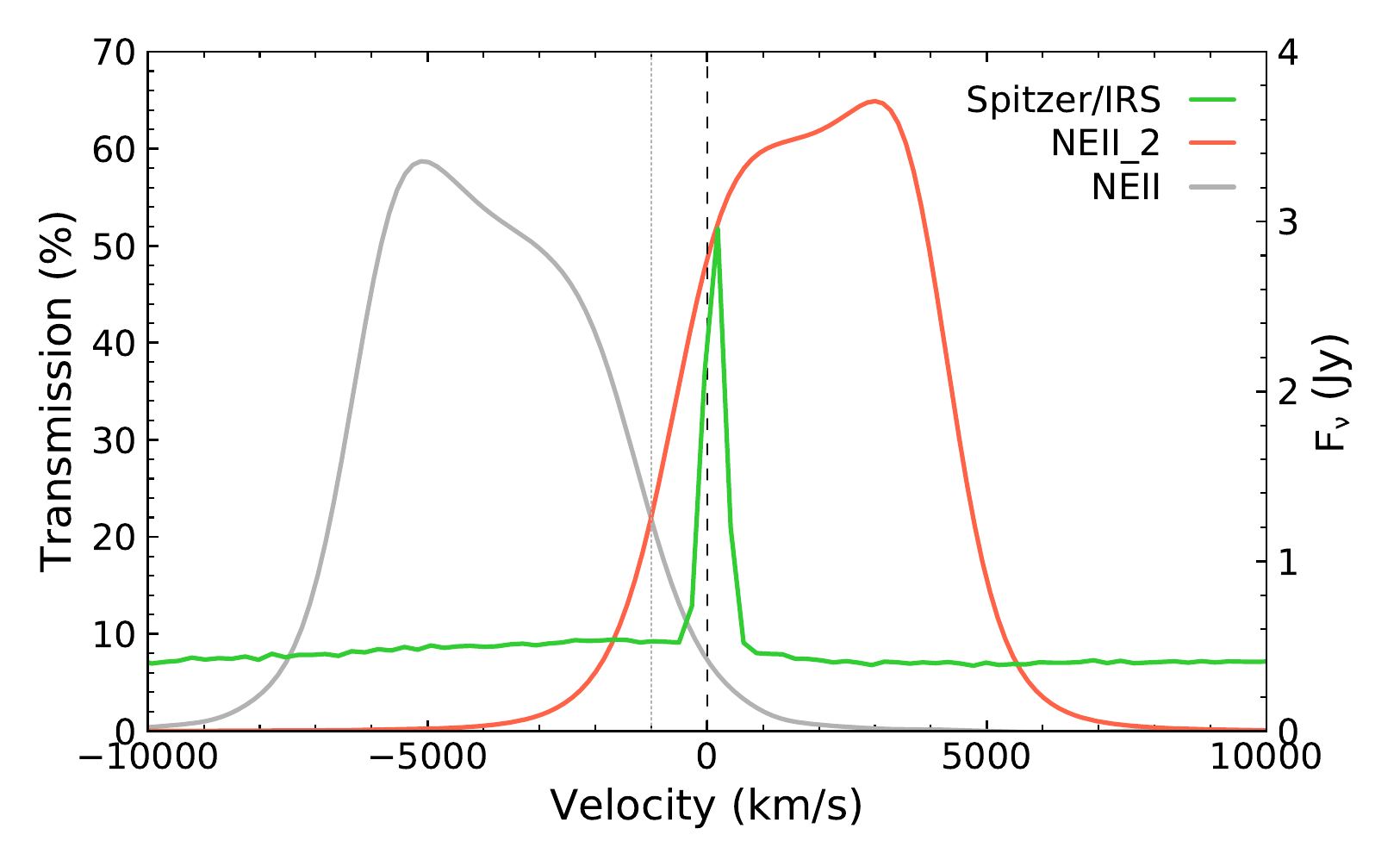}
\caption{Transmission profile of the VISIR narrow-band filters NEII (grey line) and NEII\_2 (orange line). The \textit{Spitzer}/IRS high-resolution spectrum from \citetads{2015ApJS..218...21L} suggests that contamination from PAH emission is not present within the inner kpc (green-solid line, right axis). A blueshifted emission line with $-1000\, \rm{\kms}$ (dotted vertical line) would find the same transmission in both narrow-band filters, and therefore would not appear in the continuum-subtracted map. 
}\label{fig_ne2}
\end{figure}

\clearpage

\begin{figure}
\section{Optical vs. far-IR fine-structure lines}
\centering
\includegraphics[width = 0.5\textwidth]{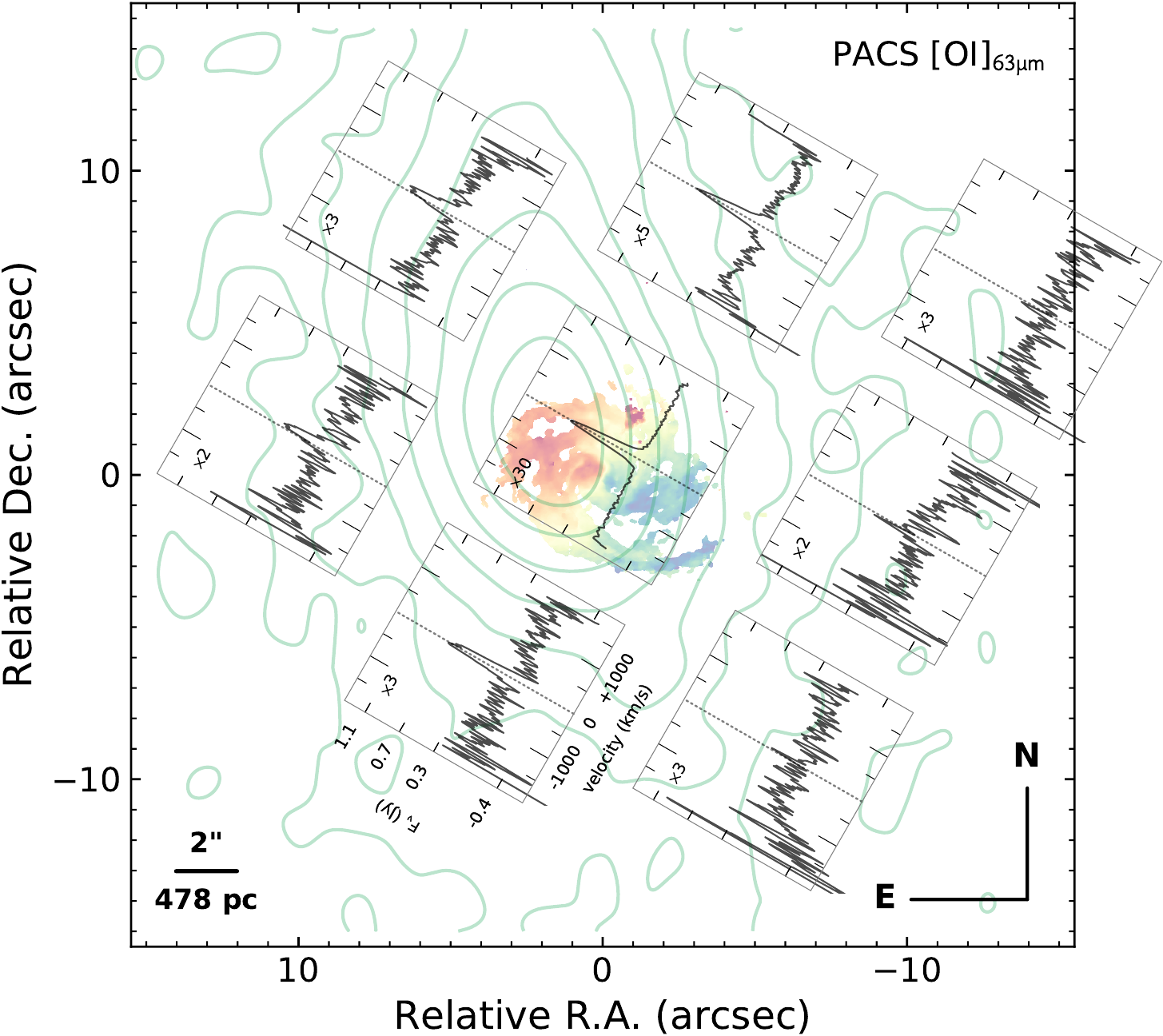}~
\includegraphics[width = 0.5\textwidth]{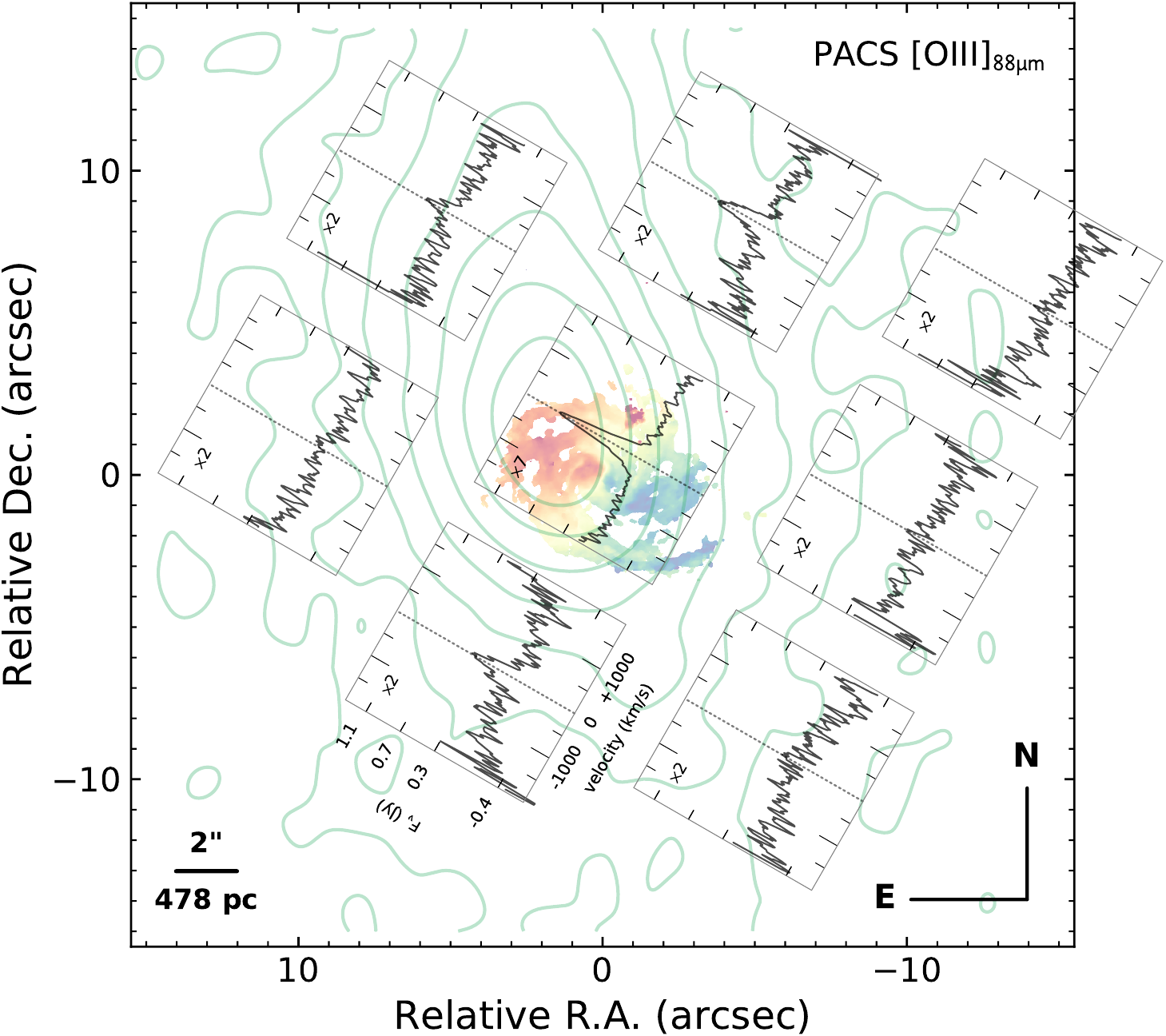}
\vspace{0.1cm}

\includegraphics[width = 0.5\textwidth]{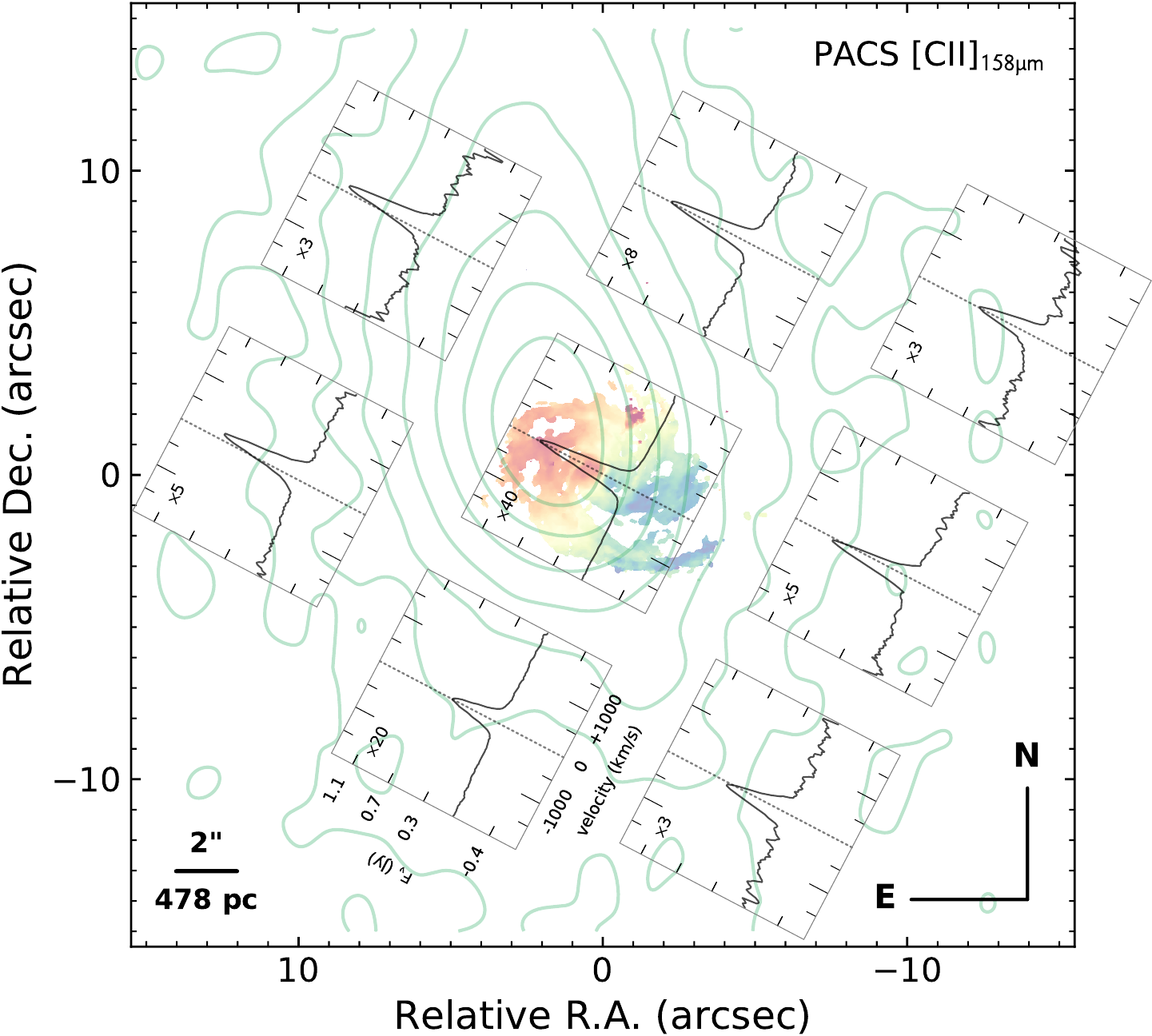}
\caption{The ALMA CO(2-1) mean velocity map is compared to the extended \oiii \ line emission (green contours), the \textit{Herschel}/PACS IFU spectra for the [\ion{O}{i}]$_{\rm 63 \mu m}$ far-IR fine-structure line (square insets in the upper-left panel), the [\ion{O}{iii}]$_{\rm 88 \mu m}$ line (upper-right panel), and the [\ion{C}{ii}]$_{\rm 158 \mu m}$ line (lower panel). The optical datacube was taken from the Siding Spring Southern Seyfert Spectroscopic Snapshot Survey \citepads{2017ApJS..232...11T} and was aligned to the ALMA data assuming the peak of the stellar optical continuum to coincide with the unresolved $1.2\, \rm{mm}$ radio core. As shown by \citetads{2017ApJS..232...11T}, the \oiii \ emission morphology is asymmetric and extended along PA\,$\sim 30\degr$. Note the lower angular resolution of the PACS data ($9\farcs4 \times 9\farcs4$ per spaxel element, only the central and neighbouring elements from the $5 \times 5$ array are shown here. [\ion{O}{i}]$_{\rm 63 \mu m}$, [\ion{O}{iii}]$_{\rm 88 \mu m}$, and [\ion{C}{ii}]$_{\rm 158 \mu m}$ fluxes are in Jy (see scale in the left inset panel of the middle row), normalised to the value indicated in the upper left corner of each frame, to ease the comparison among different frames.}\label{fig_o3_c2} 
\end{figure}

\end{appendix}

\end{document}